\documentclass[a4paper,aps,prd,twocolumn,showpacs]{revtex4-1}
\pdfoutput=1

\usepackage{enumerate}
\usepackage{graphicx}
\usepackage[caption=false]{subfig}
\usepackage{hyperref}

\usepackage{bm}
\usepackage{amsmath}
\usepackage{amssymb}

\newcommand{\Udeg}{^{\circ}}
\newcommand{\calA}{\mathcal{A}}
\newcommand{\calF}{\mathcal{F}}

\newcommand{\calP}{\mathcal{P}}
\newcommand{\ddg}{^{\,\dag\dag}}
\newcommand{\dg}{^{\,\dag}}
\newcommand{\matbar}[1]{\mat{\bar{#1}}}
\newcommand{\mat}[1]{\mathbf{#1}}
\newcommand{\ndot}[2][s]{#2^{(#1)}}
\newcommand{\ppr}{^{\,\prime\prime}}
\newcommand{\pr}{^{\,\prime}}
\newcommand{\relerr}[2]{\varepsilon(#1,#2)}
\newcommand{\trsp}{^{\mathrm{T}}}
\newcommand{\ucr}{_{\times}}
\newcommand{\upl}{_{+}}

\newcommand{\uFmetr}{_{\calF}}
\newcommand{\uess}{_{\mathrm{ess}}}
\newcommand{\ugct}{_{\mathrm{gc}}}
\newcommand{\umax}{_{\mathrm{max}}}
\newcommand{\urss}{_{\mathrm{rss}}}
\newcommand{\ussad}{_{\alpha\delta}}
\newcommand{\usslpII}{_{\mathrm{lpII}}}
\newcommand{\usslpI}{_{\mathrm{lpI}}}
\newcommand{\uss}{_{\mathrm{ss}}}
\newcommand{\utwoF}{_{0}}

\newcommand{\uX}{_{X}}
\newcommand{\uY}{_{Y}}
\newcommand{\uZ}{_{Z}}
\newcommand{\ua}{_{a}}
\newcommand{\ub}{_{b}}
\newcommand{\uc}{_{c}}
\newcommand{\uoX}{_{\mathrm{o}X}}
\newcommand{\uoY}{_{\mathrm{o}Y}}
\newcommand{\uoZ}{_{\mathrm{o}Z}}

\newcommand{\uoSigma}{_{\mathrm{o}\sigma}}
\newcommand{\uo}{_{\mathrm{o}}}
\newcommand{\usX}{_{\mathrm{s}X}}
\newcommand{\usY}{_{\mathrm{s}Y}}
\newcommand{\usZ}{_{\mathrm{s}Z}}
\newcommand{\usx}{_{\mathrm{s}x}}
\newcommand{\usy}{_{\mathrm{s}y}}
\newcommand{\usz}{_{\mathrm{s}z}}
\newcommand{\us}{_{\mathrm{s}}}
\newcommand{\ux}{_{x}}
\newcommand{\uy}{_{y}}
\newcommand{\uz}{_{z}}

\newcommand{\commitDATE}{2013-12-02 18:04:26 +0100}

\newcommand{\commitIDshort}{commitID: 90d6f2a}
\newcommand{\commitSTATUS}{CLEAN}

\begin{document}

\title{Flat parameter-space metric for all-sky searches for gravitational-wave pulsars}
\author{Karl Wette}
\email{karl.wette@aei.mpg.de}
\author{Reinhard Prix}
\email{reinhard.prix@aei.mpg.de}
\affiliation{Max-Planck-Institut f\"ur Gravitationsphysik (Albert-Einstein-Institut), D-30167 Hannover, Germany}
\affiliation{Leibniz Universit\"at Hannover, D-30167 Hannover, Germany}

\date{\commitDATE; \commitIDshort-\commitSTATUS}

\begin{abstract}
All-sky, broadband, coherent searches for gravitational-wave pulsars are computationally limited.
It is therefore important to make efficient use of available computational resources, notably by minimizing the number of templates used to cover the signal parameter space of sky position and frequency evolution.
For searches over the sky, however, the required template density (determined by the parameter-space metric) is different at each sky position, which makes it difficult in practice to achieve an efficient covering.
Previous work on this problem has found various choices of sky and frequency coordinates that render the parameter-space metric approximately constant, but which are limited to coherent integration times of either less than a few days, or greater than several months.
These limitations restrict the sensitivity achievable by hierarchical all-sky searches, and hinder the development of follow-up pipelines for interesting gravitational-wave pulsar candidates.
We present a new flat parameter-space metric approximation, and associated sky and frequency coordinates, that do not suffer from these limitations.
Furthermore, the new metric is numerically well-conditioned, which facilitates its practical use.
\end{abstract}


\maketitle

\section{Introduction}\label{sec:introduction}

Gravitational-wave pulsars are rapidly-rotating neutron stars which are hypothesized to emit continuous, narrow-band, quasi-sinusoidal gravitational waves.
Non-axisymmetric distortions of the neutron star, unstable fluid oscillations such as $r$-modes, and free precession due to misaligned symmetry and rotation axes have been proposed as possible emission mechanisms; see~\cite{Prix.2009a,Sathya.Schutz.2009} for reviews.
It remains uncertain, however, whether any of these mechanisms can generate gravitational waves strong enough to be detectable by large-scale ground-based interferometric detectors such as LIGO~\cite{Abbott.etal.2009f} or Virgo~\cite{Accadia.etal.2012a}.
Several searches using data from the first generation of these detectors have been performed; see~\cite{Abbott.etal.2010a,Abadie.etal.2012b,Aasi.etal.2013a} for recent results.
While energy-conservation-based upper limits on gravitational waves have been beaten for some individual sources~\cite{Abbott.etal.2010a,Abadie.etal.2010b,Abadie.etal.2011a}, to date no gravitational-wave pulsars have been detected.
Second-generation gravitational-wave interferometers such as Advanced LIGO~\cite{Harry.etal.2010a}, which are currently being constructed, may however be sufficiently sensitive to make a first detection~\cite{Abbott.etal.2008f,Knispel.Allen.2008a,Aasi.etal.2013a}.

The data analysis challenge of searching for gravitational-wave pulsar signals in long stretches of interferometer data is formidable.
Aside from searches for gravitational waves from known radio and X-ray pulsars, which target a single gravitational-wave template assumed to be phase-locked to the known electromagnetic signal~\cite{Abbott.etal.2010a,Abadie.etal.2011a}, searches for undiscovered gravitational-wave pulsars must cover a vast parameter space of potential signals.
For example, it is not feasible to perform a coherent search over the entire sky and a broad band of signal frequencies, despite the harnessing of $\sim10^{21}$ floating-point operations of computing power through Einstein@Home~\cite{Prix.Shaltev.2012a}, a distributed computing platform~\cite{Aasi.etal.2013a}.
This has led to the development of numerous hierarchical pipelines~\cite{Brady.etal.1998a,Krishnan.etal.2004a,Dergachev.2010a,Pletsch.2010a}, where several sensitive, computationally-expensive coherent searches of different data segments are incoherently combined using a less sensitive, but computationally cheaper, semi-coherent search.
Recent progress has been made on the optimal combination of coherent and semi-coherent searches~\cite{Cutler.etal.2005a,Prix.Shaltev.2012a}, and on the accurate estimation of the overall search sensitivity~\cite{Wette.2012a}.

A central issue in constructing a sensitive search for gravitational-wave pulsars is determining the bank of template signals to search over.
The signal template of a gravitational-wave pulsar~\cite{Jaranowski.etal.1998a} is parametrized by its sky position, often given in terms of right ascension $\alpha$ and declination $\delta$, and its frequency evolution, given most commonly by an initial frequency at some reference time, $f_0$, and a series of frequency time derivatives, or spindowns: $f_1 \equiv \dot{f}$, $f_2 \equiv \ddot{f}$, etc., up to as many as required.
The signal parameters define a manifold $\calP$ containing points $\vec\lambda = (\alpha, \delta, f_0, f_1, f_2, \dots)$, each of which corresponds to a signal template; the parameters $\vec\lambda_0$ are then coordinates in $\calP$.
The search must select a finite subset of the members of $\calP$, which in turn defines a finite bank of templates to search over.
It is improbable, however, that any real signal present in the data will possess parameters precisely matching one of the chosen templates.
At best, the real signal will be sufficiently close to one of the templates that it can be recovered with some loss in its signal-to-noise ratio.

An optimal template bank should contain a minimum number of templates, in order to reduce computational cost, with the constraint that any real signal will be recovered with some maximum acceptable loss in signal-to-noise ratio.
To achieve these constraints, the parameter space is associated with a \emph{metric}, or distance function, $g$~\cite{Balasubramanian.etal.1996a,Owen.1996a,Brady.etal.1998a}.
Given two points $\vec\lambda_0, \vec\lambda \in \calP$, the metric computes the \emph{mismatch} $\mu$, which gives the loss in signal-to-noise ratio that would result, were $\vec\lambda_0$ a real signal and $\vec\lambda$ a nearby template.
The template bank is then a finite subset of points $\{\vec\lambda_n\} \subset \calP$, such that the mismatch between any point $\vec\lambda_0 \in \calP$  and the ``closest'' template-bank member $\vec\lambda \in \{\vec\lambda_n\}$ is bounded by some prescribed maximum mismatch $\mu\umax$.
If the metric is independent of $\vec\lambda$, i.e.\ it is explicitly \emph{flat},
results from the theory of lattices can be used to place template points on a regular grid, such that the number of points required is minimized~\cite{Prix.2007b,Wette.2009a}.

The most persistent obstacle to performing optimal template placement for gravitational-wave pulsar searches has been finding a set of coordinates $\vec\lambda$, with respect to which the metric is (approximately) constant.
For searches targeting a particular point in the sky, where the search is only over the frequency evolution parameters $(f_0, f_1, f_2, \dots)$, the metric does satisfy this property~\cite{Prix.2007a,Wette.etal.2008a}, and optimal template placement was used in a search targeting the supernova remnant Cassiopeia~A~\cite{Abadie.etal.2010b}.
For searches over the sky, however, the metric is not constant with respect to the angular coordinates $(\alpha, \delta)$~\cite{Whitbeck.2006a,Prix.2007a}.
An additional practical issue, noted in \cite{Prix.2007a}, is that the metric, when expressed in conventional coordinates, is numerically highly ill-conditioned.
This makes it very difficult to, for example, compute the transformations of the metric required to implement optimal template placement.

Several alternative sky coordinates and approximate phase models have been developed, with respect to which the metric
is constant: the \emph{linear phase models} of~\cite{Jaranowski.Krolak.1999a,Astone.etal.2002b}, and the \emph{global correlation coordinates} of~\cite{Pletsch.Allen.2009a,Pletsch.2010a}.
The principal limitation of these approaches is that, in a hierarchical pipeline, the time-span of the data segments that can be coherently searched is restricted to less than a few days~\cite{Astone.etal.2002b,Pletsch.2010a}, or greater than several months~\cite{Astone.etal.2002b,Prix.2007a}.
The segment time-span is a free parameter when designing a search for gravitational-wave pulsars; one would ideally choose it based solely on trade-offs between sensitivity and computational cost, as detailed in~\cite{Prix.Shaltev.2012a}.
These restrictions, however, prevent the sensitivity of an all-sky search from being improved by increasing the length of the coherently-searched data segments beyond a few days, and it is not computationally feasible to perform an all-sky search with month-long coherent segments.
They also hinder the development of follow-up pipelines~\cite{Shaltev.Prix.2013a}, where one would like to perform more sensitive searches targeting a small number of interesting gravitational-wave pulsar candidates.

In this paper we present a new solution to these long-standing problems: an explicitly flat parameter-space metric approximation, and associated coordinates, without limitations on the coherent time-span, and where the metric is well-conditioned.
Section~\ref{sec:background} of this paper presents relevant background, and Section~\ref{sec:prior-work} examines prior research on the parameter-space metric.
Section~\ref{sec:supersky-metric} presents a new parameter-space metric approximation which is explicitly flat, but which embeds $\cal P$ in a higher-dimensional space.
Section~\ref{sec:reduced-supersky} then demonstrates how to reduce the dimensionality of the new metric back to the dimensionality of $\cal P$, while remaining constant and improving its numerical conditionedness.
Section~\ref{sec:discussion} discusses the potential uses of the new metric.
Details of the numerical simulations presented throughout this paper are found in Appendix~\ref{sec:numer-simul}.

\begin{figure}
\centering
\includegraphics[width=\linewidth]{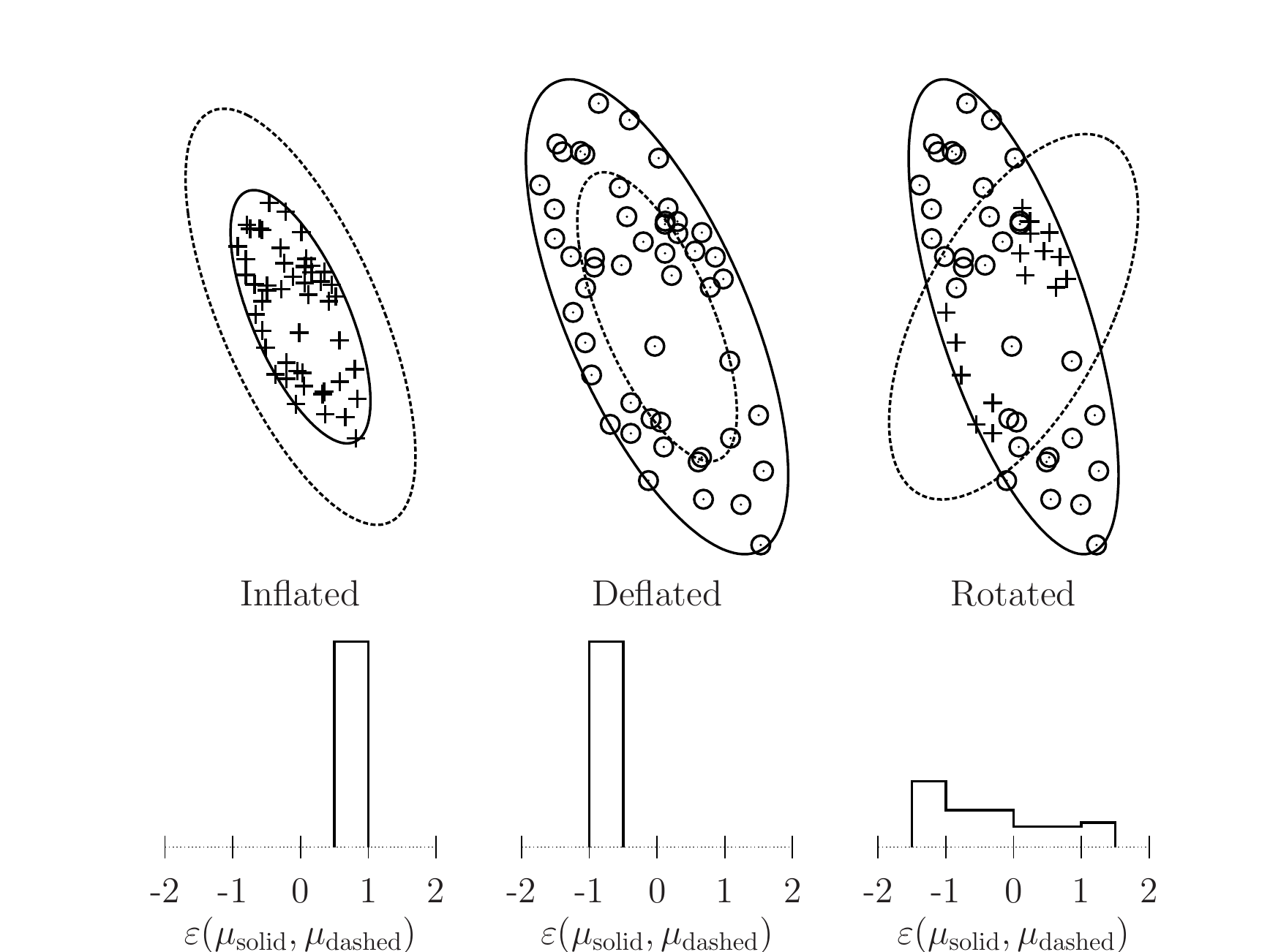}
\caption{\label{fig:relerr_cartoons}
Illustration of the behavior of the relative error $\relerr{\mu_{\mathrm{solid}}}{\mu_{\mathrm{dashed}}}$ between mismatches computed by a reference metric $\mu_{\mathrm{solid}}$, and a transformed metric $\mu_{\mathrm{dashed}}$.
Top: the reference and transformed metrics are plotted as solid and dashed ellipses.
The transformed metric has been (left to right), inflated, deflated, and rotated relative to the reference metric.
The circles and pluses represent 50 random points, sampled uniformly with respect to the reference metric; points where $\relerr{\mu_{\mathrm{solid}}}{\mu_{\mathrm{dashed}}} \le 0$ are plotted as circles, points where $\relerr{\mu_{\mathrm{solid}}}{\mu_{\mathrm{dashed}}} > 0$ are plotted as pluses.
Bottom: histograms of $\relerr{\mu_{\mathrm{solid}}}{\mu_{\mathrm{dashed}}}$ of the 50 plotted points.
}
\end{figure}

When comparing different predicted and/or calculated mismatches, $\mu_a$ and $\mu_b$, we compute their \emph{relative error}, which we define following~\cite{Prix.2007a} to be
\begin{equation}
\label{eq:relerr-def}
\relerr{\mu_a}{\mu_b} = \frac{ \mu_a - \mu_b }{ 0.5( \mu_a + \mu_b ) } \,, \quad \mu_a, \mu_b \ge 0 \,.
\end{equation}
This definition of relative error is bounded within the range $[-2, 2]$, even for large differences $|\mu_a - \mu_b| \gg 1$, while for $|\relerr{\mu_a}{\mu_b}| \ll 1$ it approaches more common definitions, e.g.\ $(\mu_a - \mu_b)/\mu_b$.
The behaviour of the relative error is illustrated in Figure~\ref{fig:relerr_cartoons}.

\section{Background}\label{sec:background}

This section presents background information relevant to this paper.
We introduce the gravitational-wave pulsar signal model (Section~\ref{sec:signal-model}), the concept of a parameter-space metric (Section~\ref{sec:metric}), and a useful approximation known as the \emph{phase metric} (Section~\ref{sec:phase-metric}).

\subsection{The signal model}\label{sec:signal-model}

A gravitational-wave pulsar signal $h(t, \calA, \vec\lambda)$, as seen in a detector, can be expressed as the sum of the products of four time-independent amplitudes, $\calA^i$, and four time-dependent basis waveforms $h_i(t, \vec\lambda)$~\cite{Jaranowski.etal.1998a}:
\begin{equation}
\label{eq:hoft-def}
h(t, \calA, \vec\lambda) = \sum_{i=1}^4 \calA^i h_i(t, \vec\lambda) \,,
\end{equation}
where $t$ is the time at the detector.
The $\calA^i$ are functions of the gravitational-wave strain amplitude $h_0$, the pulsar's angles of inclination $\iota$ and polarization $\psi$, and the wave's initial phase $\phi_0$ at a reference time $t_0$.
The $h_i(t, \vec\lambda)$ depend on the response functions $F\upl(t, \vec\lambda)$ and $F\ucr(t, \vec\lambda)$ of the detector, and on the gravitational-wave phase evolution
\begin{equation}
\label{eq:phase-ssb}
\frac{ \phi(\tau, \vec\lambda) }{2\pi} = \sum_{s=0}^{s\umax} \ndot f \frac{ (\tau - t_0)^{s+1} }{ (s+1)! } \,,
\end{equation}
where $\tau$ is the arrival time of a wavefront at the Solar System barycenter (SSB), and $\ndot f(t_0) \equiv \left. d^s f/d\tau^s\right|_{t_0}$ are the time-derivatives of the signal frequency $f(\tau)$ at the SSB.
The gravitational-wave phase at the detector is found by substituting
\begin{equation}
\label{eq:tau-to-t}
\tau(t, \vec\lambda) - t_0 = (t - t_0) + \frac{ \vec r(t) \cdot \vec n }{c} + \Delta_{\mathrm{relativistic}} \,,
\end{equation}
where $t$ is the arrival time of the wavefront at the detector, $\vec r(t)$ is the detector position vector relative to the SSB, $\vec n$ is a unit vector pointing from the SSB to the pulsar's position in the sky, and $\Delta_{\mathrm{relativistic}}$ represents the relativistic Einstein and Shapiro delays.
The second term of Eq.~\eqref{eq:tau-to-t} is also known as the R\o{}mer delay.

The result of coherently matched-filtering the signal model with detector data and maximizing over the unknown amplitudes $\calA^i$ is known in this context as the $\calF$-statistic~\cite{Jaranowski.etal.1998a,Cutler.Schutz.2005a}.
For a signal with parameters $(\calA^i, \vec\lambda_0)$, the $\calF$-statistic in a template $\vec\lambda$ follows a non-central $\chi^2$-distribution with four degrees of freedom and a non-centrality parameter given by the squared signal-to-noise ratio (SNR) $\rho^2(\calA, \vec\lambda_0; \vec\lambda)$.
For perfectly-matched signals,
\begin{equation}
\label{eq:snr-def}
\rho^2(\calA, \vec\lambda_0; \vec\lambda_0)  = \frac{2}{S_h(f_0)} \int_{t_0-T/2}^{t_0+T/2} dt \, h^2(t, \calA, \vec\lambda_0) \,,
\end{equation}
where $S_h(f_0)$ is the (single-sided) power spectral density (PSD) of the detector noise at the signal frequency $f_0$, and $T$ is the time spanned by the coherently-analyzed data.
For simplicity, in this paper we will assume that the detector data is continuous (i.e.\ contains no gaps), and that the PSD in a sufficiently small frequency band surrounding a signal is constant in time and frequency.
These limitations are readily addressed in a real implementation of the $\calF$-statistic~\cite{Prix.2010a}.

\subsection{The metric}\label{sec:metric}

The mismatch $\mu$ between a signal with parameters $\vec\lambda_0$ and a nearby template with parameters $\vec\lambda$ is defined in terms of the squared SNR~\cite{Brady.etal.1998a,Prix.2007a}:
\begin{equation}
\label{eq:Fstat-mismatch}
\mu\utwoF = \frac{ \rho^2(\calA, \vec\lambda_0; \vec\lambda_0) - \rho^2(\calA, \vec\lambda_0; \vec\lambda) }{ \rho^2(\calA, \vec\lambda_0; \vec\lambda_0) } \,,
\end{equation}
where the mismatched SNR $\rho^2(\calA, \vec\lambda_0; \vec\lambda)$ is given by Eq.~(28) of~\cite{Prix.2007a}.
In the numerical simulations presented in this paper, $\mu\utwoF$ is calculated as follows: a gravitational-wave pulsar signal is generated with parameters $\vec\lambda_0$, and searched for using the $\calF$-statistic at points $\vec\lambda_0$ and $\vec\lambda$, returning the values $\calF(\vec\lambda_0)$ and $\calF(\vec\lambda)$ respectively.
The mismatch is then calculated using Eq.~\eqref{eq:Fstat-mismatch} and the relation
\begin{equation}
\label{eq:E2F-to-SNR}
2\calF(\vec\lambda) = E[2\calF(\vec\lambda)] = 4 + \rho^2(\calA, \vec\lambda_0; \vec\lambda) \,.
\end{equation}
No simulated noise is added to the gravitational-wave pulsar signal, and thus $2\calF(\vec\lambda)$ is equal to its expectation value $E[2\calF(\vec\lambda)]$.

If the difference $\Delta\vec\lambda = \vec\lambda - \vec\lambda_0$ is small enough, $\rho^2(\calA, \vec\lambda_0; \vec\lambda)$ can be Taylor-expanded with respect to $\rho^2(\calA, \vec\lambda_0; \vec\lambda_0)$.
The $\calF$-statistic mismatch $\mu\utwoF$ is then approximated by
\begin{equation}
\label{eq:mismatch-metr}
\mu\utwoF \approx \mu_{\mat g} \equiv \Delta\vec\lambda \cdot \mat g \Delta\vec\lambda \,.
\end{equation}
The metric mismatch $\mu_{\mat g}$ is calculated via the metric $\mat g$, whose coefficients are
\begin{equation}
\label{eq:metric-coeff-def}
g(\lambda_i, \lambda_j) = \frac{-1}{2 \rho^2(\calA, \vec\lambda_0; \vec\lambda_0)} \left. \frac{\partial \rho^2(\calA, \vec\lambda_0; \vec\lambda)}{\partial \lambda_i \partial \lambda_j} \right|_{\vec\lambda = \vec\lambda_0} \,.
\end{equation}
There are no terms proportional to the first derivatives of $\rho^2(\calA, \vec\lambda_0; \vec\lambda)$ with respect to $\vec\lambda$, since by definition $\rho^2(\calA, \vec\lambda_0; \vec\lambda)$ is a maximum at the signal location $\vec\lambda_0$.

The matrix $\mat g$ is positive definite by construction~\cite{Prix.2007a}, and thus the region $\mu \leq \mu\umax$ forms an ellipsoid, centered on $\vec\lambda_0$, in the parameter space $\calP$.
If $\mat g$ is flat, each template point $\vec\lambda_n \in \calP$ will be surrounded by an identical ellipsoid.
We can then apply a global coordinate transformation to $\calP$ which maps the ellipsoids to spheres, each with a template point at its center.
The problem of template placement is now equivalent to the sphere-covering problem in lattice theory~\cite{Prix.2007b}, and the solution which minimizes the number of template points is to place them at the vertices of a lattice which is known to achieve the best possible covering.
The best choice of lattice depends on the dimensionality of $\calP$; for example, in 2 dimensions it is the hexagonal lattice~\cite{Conway.Sloane.1988}.
If $\mat g$ is not flat (or just non-constant), however, other methods of template placement, such as random or stochastic algorithms~\cite{Messenger.etal.2009a,Harry.etal.2009a,Manca.Vallisneri.2010a}, must be employed.

\subsection{The phase metric}\label{sec:phase-metric}

For the $\calF$-statistic, the $g(\lambda_i, \lambda_j)$ are complicated functions depending on the unknown amplitudes $\calA$, as described in~\cite{Prix.2007a}, which include both derivatives of the amplitude modulation $F\upl(t,\vec\lambda)$ and $F\ucr(t,\vec\lambda)$ of the signal, and derivatives of the phase modulation, given by $\phi(t,\vec\lambda)$.
If $T$ is large compared to a day, however, the contribution of the more rapid ($\gtrsim 100$/s) phase modulation dominates that of the slower ($\lesssim 1$/day) amplitude modulation.
In this limit, the metric reduces to a simplified form, known as the phase metric, whose coefficients involve only derivatives of $\phi(t,\vec\lambda)$:
\begin{equation}
\label{eq:phase-metric-def}
g(\lambda_i, \lambda_j) =
\left[ \frac{\partial \phi(t,\vec\lambda)}{\partial \lambda_i}, \frac{\partial \phi(t,\vec\lambda)}{\partial \lambda_j} \right] \,,
\end{equation}
where we define the operators
\begin{align}
\label{eq:phase-metric-ops}
\big[x(t), y(t)\big] &= \big\langle x(t) y(t) \big\rangle - \big\langle x(t) \big\rangle \big\langle y(t) \big\rangle \,, \\
\big\langle x(t) \big\rangle &= \frac{1}{T} \int_{t_0-T/2}^{t_0+T/2} dt \, x(t) \,.
\end{align}
An equivalent expression for the metric was also obtained in~\cite{Brady.etal.1998a} by instead assuming a simplified signal model where the amplitude motion is discarded.
If the phase $\phi(t,\vec\lambda)$ is linear in the coordinates $\vec\lambda$, then the $g(\lambda_i, \lambda_j)$ are independent of
$\vec\lambda$, and $\mat g$ is therefore flat.
Thus, the problem of finding a constant metric approximation is reduced to one of linearizing $\phi(t,\vec\lambda)$ with respect to its coordinates.

To linearize $\phi(t,\vec\lambda)$ with respect to the frequency and spindown coordinates $\ndot f$, we first substitute Eq.~\eqref{eq:tau-to-t} into Eq.~\eqref{eq:phase-ssb}, neglecting the relativistic terms which are not important for template placement:
\begin{equation}
\label{eq:phase-det-subst}
\frac{ \phi(t, \vec\lambda) }{2\pi} \approx \sum_{s=0}^{s\umax} \frac{ \ndot f }{ (s+1)! }
\left[ \Delta t + \frac{ \vec r(t) \cdot \vec n }{c} \right]^{s+1} \,,
\end{equation}
where $\Delta t = t - t_0$.
We now expand the factor $[\dots]^{s+1}$ and retain only the first two leading order terms in $\Delta t$.
This approximation can be made because $\vec r(t) \cdot \vec n / c \lesssim 500$~seconds (the approximate light travel time from the Sun to the Earth) while $\Delta t \sim T$, and so $\Delta t \gg \vec r(t) \cdot \vec n / c$ for $T \gtrsim$~days.
The approximate Eq.~\eqref{eq:phase-det-subst} now reads:
\begin{equation}
\label{eq:phase-det-approx}
\frac{ \phi(t, \vec\lambda) }{2\pi} \approx \sum_{s=0}^{s\umax} \ndot f \frac{ \Delta t^{s+1} }{ (s+1)! }
+ \frac{ \vec r(t) \cdot \vec n }{c} \sum_{s=0}^{s\umax} \ndot f \frac{ \Delta t^s }{ s! } \,.
\end{equation}
The summation in the second right-hand-side term is precisely $f(t)$, the instantaneous frequency of the signal at time $t$.
Using the same argument, we see that the derivatives of $\phi(t, \vec\lambda)$ with respect to the $\ndot f$ [which will appear in Eq.~\eqref{eq:phase-metric-def}] of the second term will be small relative to the first, i.e.\ $\Delta t^{s+1} \gg \Delta t^{s} \vec r(t) \cdot \vec n / c$.
Hence, $f(t)$ may be approximated by some constant $f\umax$, usually chosen conservatively to be the maximum of $f(t)$ over $T$.
The resulting approximate phase is now:
\begin{equation}
\label{eq:phase-det}
\frac{ \phi(t, \vec\lambda) }{2\pi} \approx \sum_{s=0}^{s\umax} \ndot f \frac{ \Delta t^{s+1} }{ (s+1)! }
+ \frac{ \vec r(t) \cdot \vec n }{c} f\umax \,.
\end{equation}

\section{Prior work}\label{sec:prior-work}

While it is straightforward to obtain a linear phase model $\phi(t, \vec\lambda)$ with respect to the frequency and spindown coordinates $\ndot f$, as shown in Section~\ref{sec:phase-metric}, the same cannot be said of the sky coordinates, which enter Eq.~\eqref{eq:phase-ssb} through the sky position vector $\vec n$.
If, for example, we choose right ascension $\alpha$ and declination $\delta$ as sky coordinates, then the derivatives of $\phi(t, \vec\lambda)$ themselves depend on the sky coordinates:
\begin{equation}
\label{eq:phi-deriv-alpha-delta}
\begin{split}
\mathrm{d}\phi &\propto ( r\uz \cos\delta - r\uy \sin\alpha \sin\delta - r\ux \cos\alpha \sin\delta ) \mathrm{d}\delta \\
&\quad + ( r\uy \cos\alpha \cos\delta - r\ux\sin\alpha \cos\delta ) \mathrm{d}\alpha + \dots \,,
\end{split}
\end{equation}
where $\vec n = (\cos\alpha \cos\delta, \sin\alpha \cos\delta, \sin\delta)$, and $\vec r(t) = (r\ux, r\uy, r\uz)$ are expressed in equatorial coordinates $(x,y,z)$.
If, instead, two components of the vector $\vec n = (n\ux, n\uy, n\uz)$ are chosen, e.g.\ $n\ux$ and $n\uy$, the constraint $|\vec n| = 1$ requires that the third component is a function of the other two, i.e.\ $n\uz = (1 - n\ux^2 - n\uy^2)^{1/2}$, and so the derivatives of $\phi(t, \vec\lambda)$ still depend on the coordinates:
\begin{equation}
\label{eq:phi-deriv-nxy}
\mathrm{d}\phi \propto \left( r\ux - r\uz \frac{n\ux}{n\uz} \right) \mathrm{d}n\ux +
\left( r\uy - r\uz \frac{n\uy}{n\uz} \right) \mathrm{d}n\uy + \dots \,.
\end{equation}

This section presents two prior approaches to this problem: the linear phase models (Section~\ref{sec:linear-phase-models}), and the global correlation coordinates (Section~\ref{sec:glob-corr-coord}).
The new approach to this problem taken in this paper is presented in Section~\ref{sec:supersky-metric}.

\subsection{Linear phase models}\label{sec:linear-phase-models}

\begin{figure*}
\centering
\subfloat[]{\includegraphics[width=0.49\linewidth]{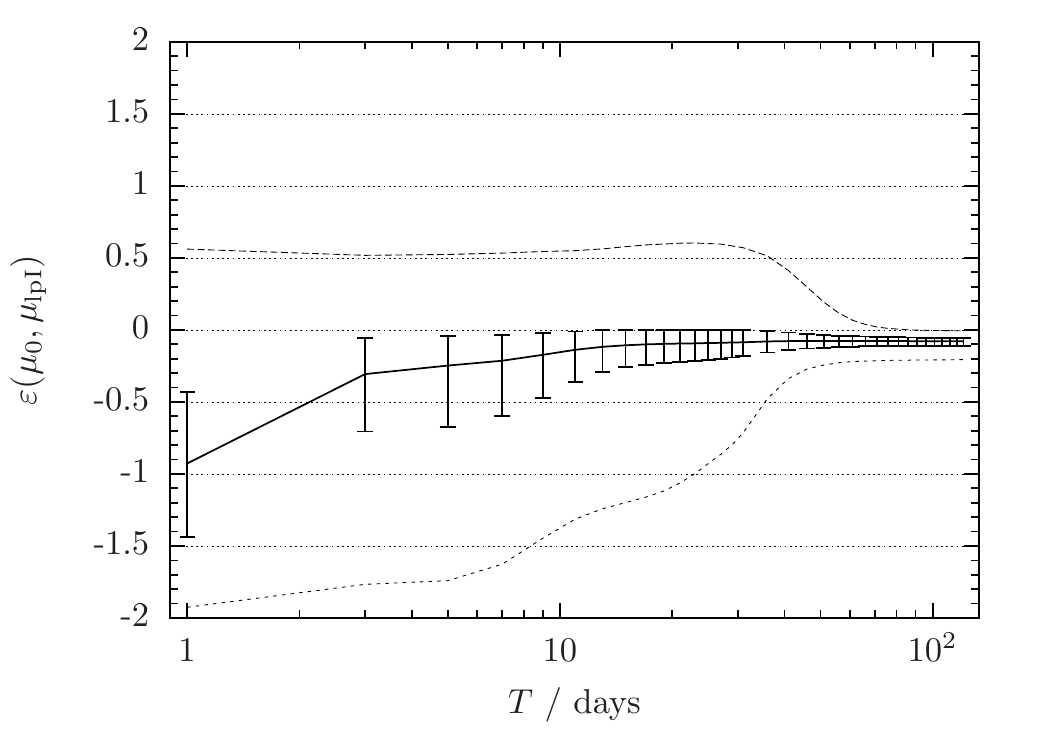}\label{fig:mu_re_f1dot_H1_twoF_sslpI_mu0p2}}
\subfloat[]{\includegraphics[width=0.49\linewidth]{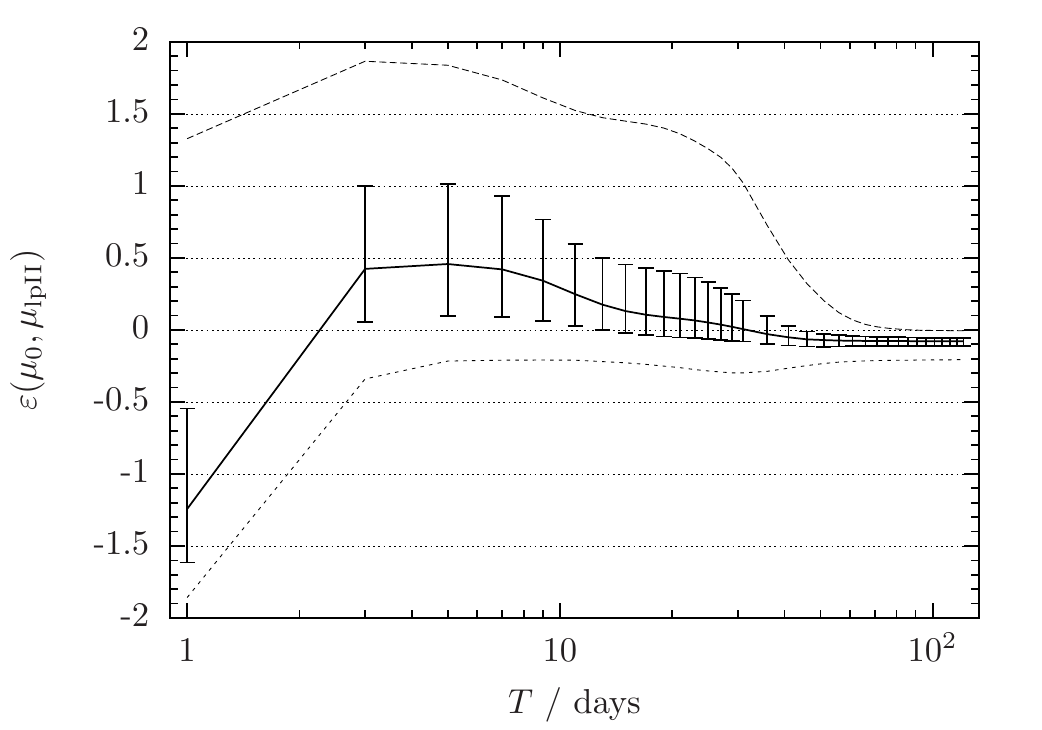}\label{fig:mu_re_f1dot_H1_twoF_sslpII_mu0p2}}\\
\subfloat[]{\includegraphics[width=0.49\linewidth]{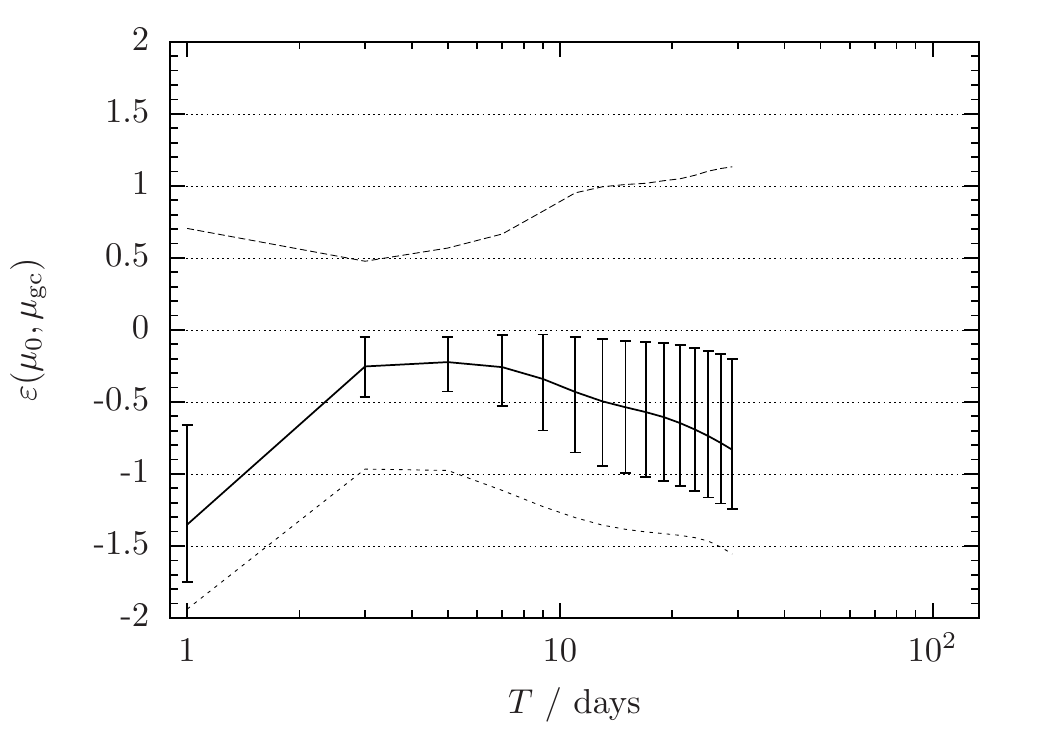}\label{fig:mu_re_f1dot_H1_twoF_gct_mu0p2}}
\subfloat[]{\includegraphics[width=0.49\linewidth]{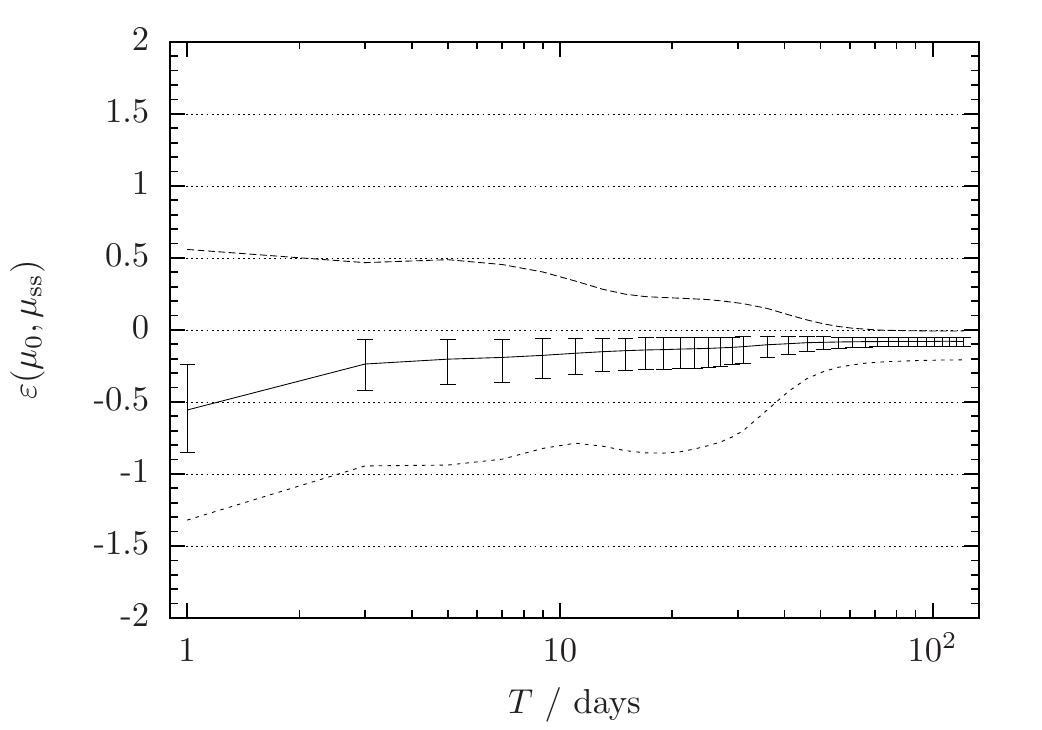}\label{fig:mu_re_f1dot_H1_twoF_ss_mu0p2}}\\
\subfloat[]{\includegraphics[width=0.49\linewidth]{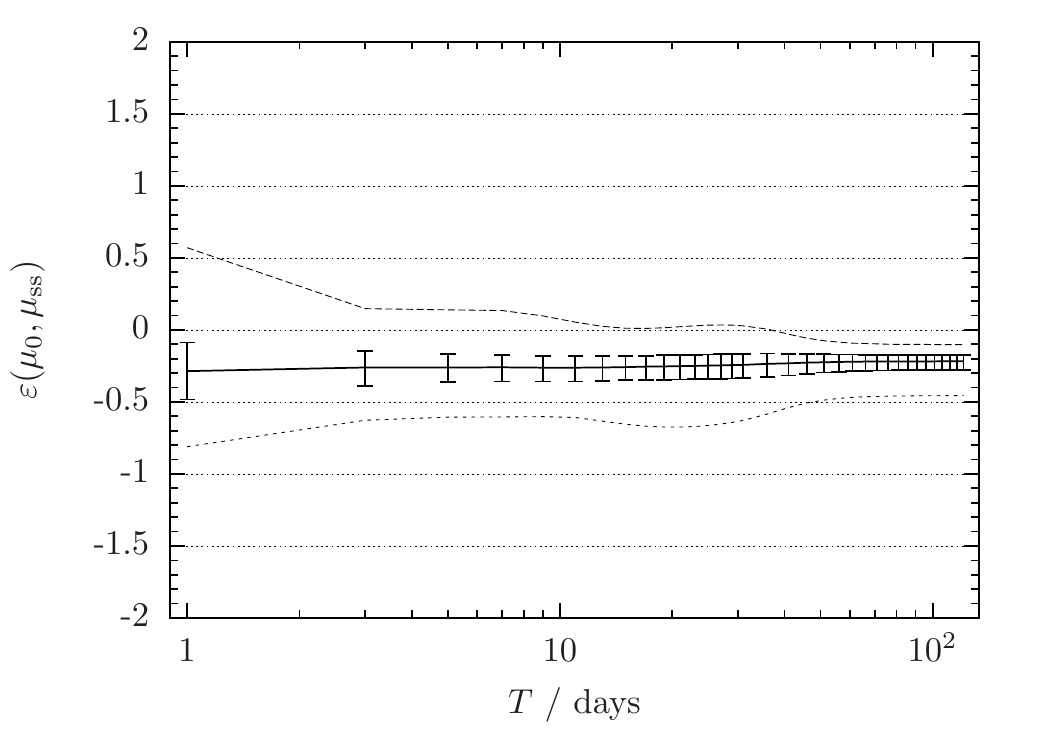}\label{fig:mu_re_f1dot_H1_twoF_ss_mu0p6}}
\subfloat[]{\includegraphics[width=0.49\linewidth]{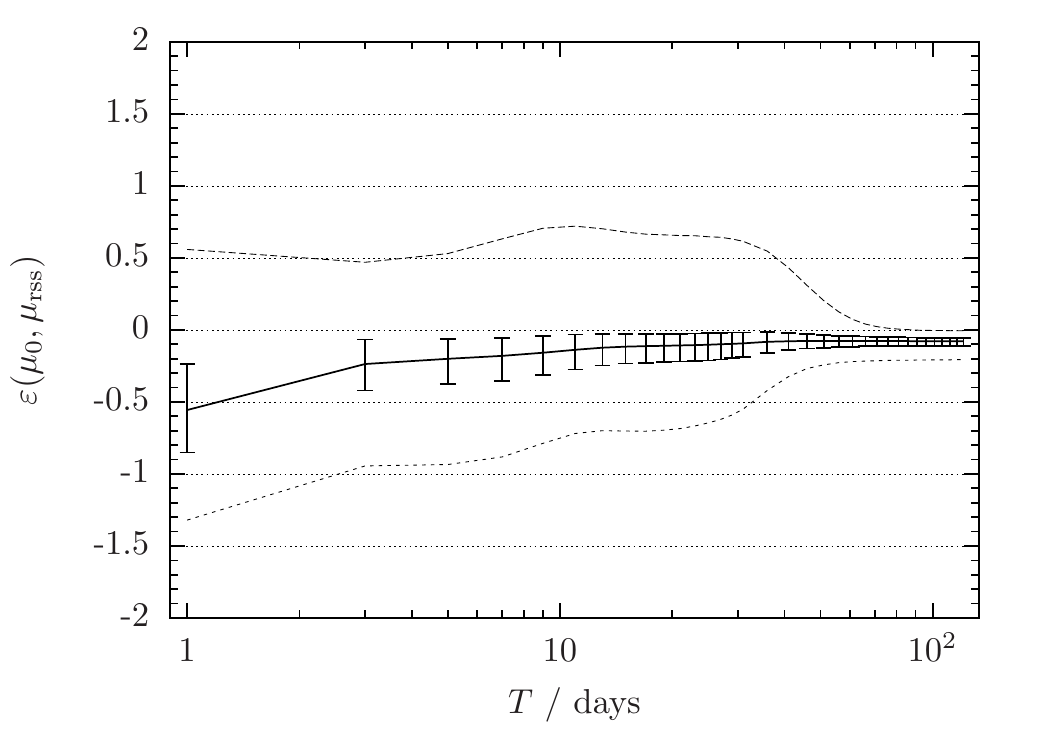}\label{fig:mu_re_f1dot_H1_twoF_rss_mu0p2}}
\caption{\label{fig:mu_re_f1dot_H1_twoF}
Relative errors as a function of $T$ between mismatches calculated from the $\calF$-statistic, $\mu\utwoF$, and predicted by:
\protect\subref{fig:mu_re_f1dot_H1_twoF_sslpI_mu0p2} the phase metric using linear phase model I, $\relerr{\mu\utwoF}{\mu\usslpI}$, for $\mu\utwoF \le 0.2$;
\protect\subref{fig:mu_re_f1dot_H1_twoF_sslpII_mu0p2} the phase metric using linear phase model II, $\relerr{\mu\utwoF}{\mu\usslpII}$, for $\mu\utwoF \le 0.2$;
\protect\subref{fig:mu_re_f1dot_H1_twoF_gct_mu0p2} the global correlation metric, $\relerr{\mu\utwoF}{\mu\ugct}$, for $\mu\utwoF \le 0.2$ and $T < 30$~days;
\protect\subref{fig:mu_re_f1dot_H1_twoF_ss_mu0p2},~\protect\subref{fig:mu_re_f1dot_H1_twoF_ss_mu0p6} the supersky metric, $\relerr{\mu\utwoF}{\mu\uss}$, for $\mu\utwoF \le 0.2$ and $0.2 \le \mu\utwoF \le 0.6$ respectively;
and \protect\subref{fig:mu_re_f1dot_H1_twoF_rss_mu0p2} the reduced supersky metric, $\relerr{\mu\utwoF}{\mu\urss}$, for $\mu\utwoF \le 0.2$.
Plotted are the median (solid line), the 25th--75th percentile range (error bars), and the 2.5th (dotted line) and 97.5th (dashed line) percentiles.
Only first spindown is used.
}
\end{figure*}

The linear phase models of~\cite{Jaranowski.Krolak.1999a,Astone.etal.2002b} express $\vec n = (n\uX, n\uY, n\uZ)$ in ecliptic coordinates $(X,Y,Z)$, and adopt the $X$ and $Y$ components $(n\uX, n\uY)$ as sky coordinates.
The restriction $|\vec n| = 1$ then requires $n\uZ = \sqrt{1 - n\uX^2 - n\uY^2}$.
We write
\begin{align}
\label{eq:r-dot-n-lpm}
\vec r(t) \cdot \vec n &= r\uX(t) \, n\uX + r\uY(t) \, n\uY + r\uZ(t) \, n\uZ \\
\begin{split}
&= [r\usX(t) + r\uoX(t)] \, n\uX \\
&\qquad + [r\usY(t) + r\uoY(t)] \, n\uY + r\usZ(t) \, n\uZ \,,
\end{split}
\end{align}
where the detector position vector $\vec r(t) = \vec r\us(t) + \vec r\uo(t)$ is decomposed into its diurnal and orbital components in ecliptic coordinates, $\vec r\us(t) = [r\usX(t), r\usY(t), r\usZ(t)]$ and $\vec r\uo(t) = [r\uoX(t), r\uoY(t), 0]$ respectively.
(This assumes a planar Earth orbit, which is not exactly satisfied in reality, due to e.g. the Earth--Moon interaction.)
We see that $\vec r(t) \cdot \vec n$ is linear in $n\uX$ and $n\uY$ only if $r\usZ(t) = 0$, i.e.\ only if the diurnal motion of the detector in the ecliptic $Z$ direction is neglected.
Two linear phase models are presented in~\cite{Jaranowski.Krolak.1999a} which achieve this: in linear phase model I, $r\usZ(t)$ is discarded; in linear phase model II (also known as the \emph{orbital metric}~\cite{Prix.2007a}), the entire diurnal motion $\vec r\us(t)$ is discarded.
Simulations investigating the accuracy of signal parameter estimation using the linear phase models were presented in~\cite{Jaranowski.Krolak.1999a}.

Figures~\ref{fig:mu_re_f1dot_H1_twoF_sslpI_mu0p2} and~\ref{fig:mu_re_f1dot_H1_twoF_sslpII_mu0p2} show the relative errors $\relerr{\mu\utwoF}{\mu\usslpI}$ and $\relerr{\mu\utwoF}{\mu\usslpII}$ between mismatches predicted by the phase metric using the linear phase models I and II versus the mismatch $\mu\utwoF$ calculated from the
$\calF$-statistic, as given in Eq.~\eqref{eq:Fstat-mismatch}.
Linear phase model I (Figure~\ref{fig:mu_re_f1dot_H1_twoF_sslpI_mu0p2}) models the $\calF$-statistic mismatch well, converging to $|\relerr{\mu\utwoF}{\mu\usslpI}| \lesssim 0.1$ for $T \gtrsim 20$~days.
The spread of errors can, however, be quite large: the 25th--75th percentile range (error bars) is $\gtrsim 0.5$ for $T \lesssim 10$~days; the 2.5th--97.5th percentile range (dotted to dashed lines) is $\gtrsim 1$ for $T \lesssim 35$~days.
Linear phase model II (Figure~\ref{fig:mu_re_f1dot_H1_twoF_sslpII_mu0p2}) is also a reasonable approximation in the longer-$T$ limit, consistent with similar simulations presented in~\cite{Prix.2007a}.
It initially under-estimates $\mu\utwoF$ [i.e.\ $\relerr{\mu\utwoF}{\mu\usslpII} > 0$], up to 0.5 at $T \sim 5$~days, before converging to an over-estimate [$\relerr{\mu\utwoF}{\mu\usslpII} < 0$] of $\sim 0.1$ for $T \gtrsim 30$~days.

The relative errors between the linear phase models and the $\calF$-statistic exhibit features common to most of the phase metrics examined in this paper.
In general, the phase metrics tend to over-estimate the $\calF$-statistic mismatch [i.e.\ $\relerr{\mu\utwoF}{\mu_{\cdots}} < 0$], which would lead to a conservative, over-dense template bank.
This is due to the neglection of higher-order terms in $\Delta\vec\lambda$ in the derivation of the metric [Eq.~\eqref{eq:mismatch-metr}], which leads to the metric-predicted mismatch typically being larger than the $\calF$-statistic mismatch for the same $\Delta\vec\lambda$.
This effect prevents the phase metric from exactly predicting the $\calF$-statistic mismatch at long $T$.
In addition, at $T \sim 1$~day, the diurnal amplitude modulation of the gravitational-wave pulsar signal, neglected in the phase metric approximation, changes the size and orientation of the full $\calF$-statistic metric~\cite{Prix.2007a} relative to the phase metric.
As illustrated in Figure~\ref{fig:relerr_cartoons}, this can lead to further over-estimation of the $\calF$-statistic mismatch by the phase metric.

\subsection{Global correlation coordinates}\label{sec:glob-corr-coord}

\begin{figure*}
\centering
\includegraphics[width=\linewidth]{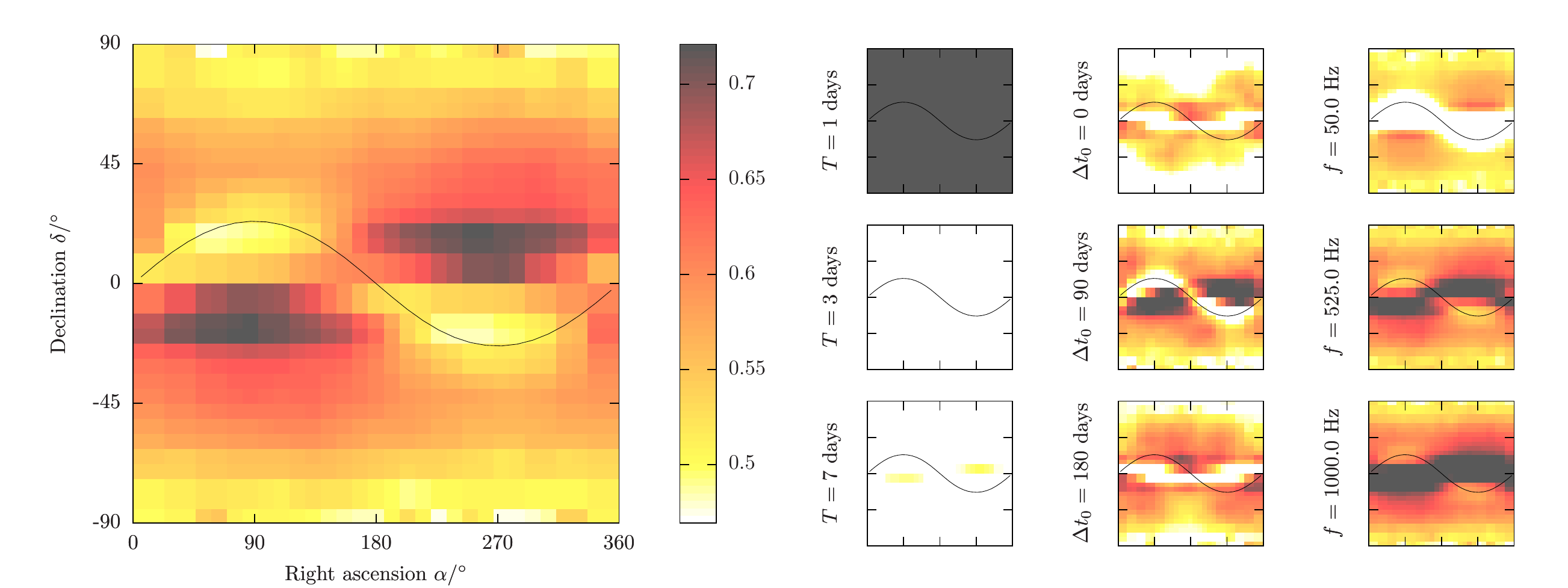}
\caption{\label{fig:mu_sm_f1dot_H1_twoF_gct_mu0p2}
Median magnitude of the relative error $|\relerr{\mu\utwoF}{\mu\ugct}|$, as a function of $\alpha$ and $\delta$, for $\mu\utwoF \le 0.2$.
The ecliptic equator is over-plotted in black.
Left: median $|\relerr{\mu\utwoF}{\mu\ugct}|$ over time-spans $1 \le T \le 29$~days, and over all simulation reference times $\Delta t_0$ and maximum frequencies $f\umax$; see Appendix~\ref{sec:numer-simul}.
Right: median $|\relerr{\mu\utwoF}{\mu\ugct}|$ at fixed values of $T$, $\Delta t_0$, and $f\umax$; axis ranges and color values are the same as for the left-hand-side plot.
Only first spindown is used.
}
\end{figure*}

The global correlation coordinates of~\cite{Pletsch.Allen.2009a,Pletsch.2010a} adopt the $x$ and $y$ coefficients of $\vec n$ in equatorial coordinates, $(n\ux, n\uy)$, as sky coordinates.
Decomposing the detector position vector into its diurnal and orbital components in equatorial coordinates, $\vec r\us(t) = [r\usx(t), r\usy(t), 0]$ and $\vec r\uo(t)$ respectively, we have
\begin{equation}
\label{eq:r-dot-n-gct}
\vec r(t) \cdot \vec n = r\usx(t) n\ux + r\usy(t) n\uy + \vec r\uo(t) \cdot \vec n \,.
\end{equation}
The global correlation coordinates then absorb $\vec r\uo(t) \cdot \vec n$ into the frequency and spindown parameters $\ndot f$, by Taylor-expanding $\vec r\uo(t)$ in time, and defining new parameters $\ndot \nu(t)$ which are functions of $\ndot f$ and $\vec n$.
Assuming that no second- or higher-order spindowns are required, i.e.\ $\ndot f = 0$ for $s > 1$, the global correlation coordinates $\nu(t)$ at time $t$ are functions of the frequency $f(t) = f + \dot f \Delta t$ and spindown $\dot f$ at time $t$~\cite{Pletsch.Allen.2009a}:
\begin{subequations}
\label{eqs:gct-nu}
\begin{align}
\nu(t) &= f(t) + \frac{ f(t) \dot{\vec r}\uo(t) + \dot f \vec r\uo(t) }{c} \cdot \vec n \,, \\
\dot \nu(t) &= \dot f + \frac{ f(t) \ddot{\vec r}\uo(t) + 2 \dot f \dot{\vec r}\uo(t) }{c} \cdot \vec n \,.
\end{align}
\end{subequations}
The phase is then linear in the coordinates $n\ux$, $n\uy$, $\nu(t)$, and $\dot \nu(t)$.
The approximation of $\vec r\uo(t)$ as a Taylor series limits the validity of these coordinates to $T \lesssim 2$--10 days, depending on the frequency searched~\cite{Pletsch.2010a}.
A similar linearized phase model is also presented in~\cite{Astone.etal.2002b}; they found that the adequacy of this model was limited to $T \lesssim 8$--14 days, depending on the search frequency, number of spindowns, and the parameter estimation accuracy required.
An examination of the limitations of the global correlation method is presented in~\cite{Manca.Prix.2013}.

Figure~\ref{fig:mu_re_f1dot_H1_twoF_gct_mu0p2} plots the relative error $\relerr{\mu\utwoF}{\mu\ugct}$ between mismatches predicted by the global correlation metric, and calculated from the $\calF$-statistic via Eq.~\eqref{eq:Fstat-mismatch}, up to $T < 30$~days.
The global correlation coordinates perform best when $3 \lesssim T \lesssim 7$~days; below $\sim 3$~days and above $\sim 15$~days, $|\relerr{\mu\utwoF}{\mu\ugct}| \gtrsim 0.5$, broadly consistent with the domain of validity found in~\cite{Pletsch.2010a}.
Unlike other phase metrics examined in this paper, the global correlation coordinates perform worse at long $T$, due to the breakdown of the Taylor expansion of the orbital motion.
Figure~\ref{fig:mu_sm_f1dot_H1_twoF_gct_mu0p2} plots the median error magnitude $|\relerr{\mu\utwoF}{\mu\ugct}|$ as a function of sky position, over the full ranges of simulation parameters, and at fixed values of $T$, $\Delta t_0$, and $f\umax$.
The smallest $|\relerr{\mu\utwoF}{\mu\ugct}|$ are along the ecliptic equator and at the poles; the largest are at the points $\alpha = 180\Udeg \pm 90\Udeg$, $\delta = \pm 20\Udeg$.
The errors are independent of sky position at fixed $T$, but become sky position-dependent when considering fixed $\Delta t_0$ and $f\umax$.

\section{The supersky metric}\label{sec:supersky-metric}

\begin{figure}
\includegraphics[width=\linewidth]{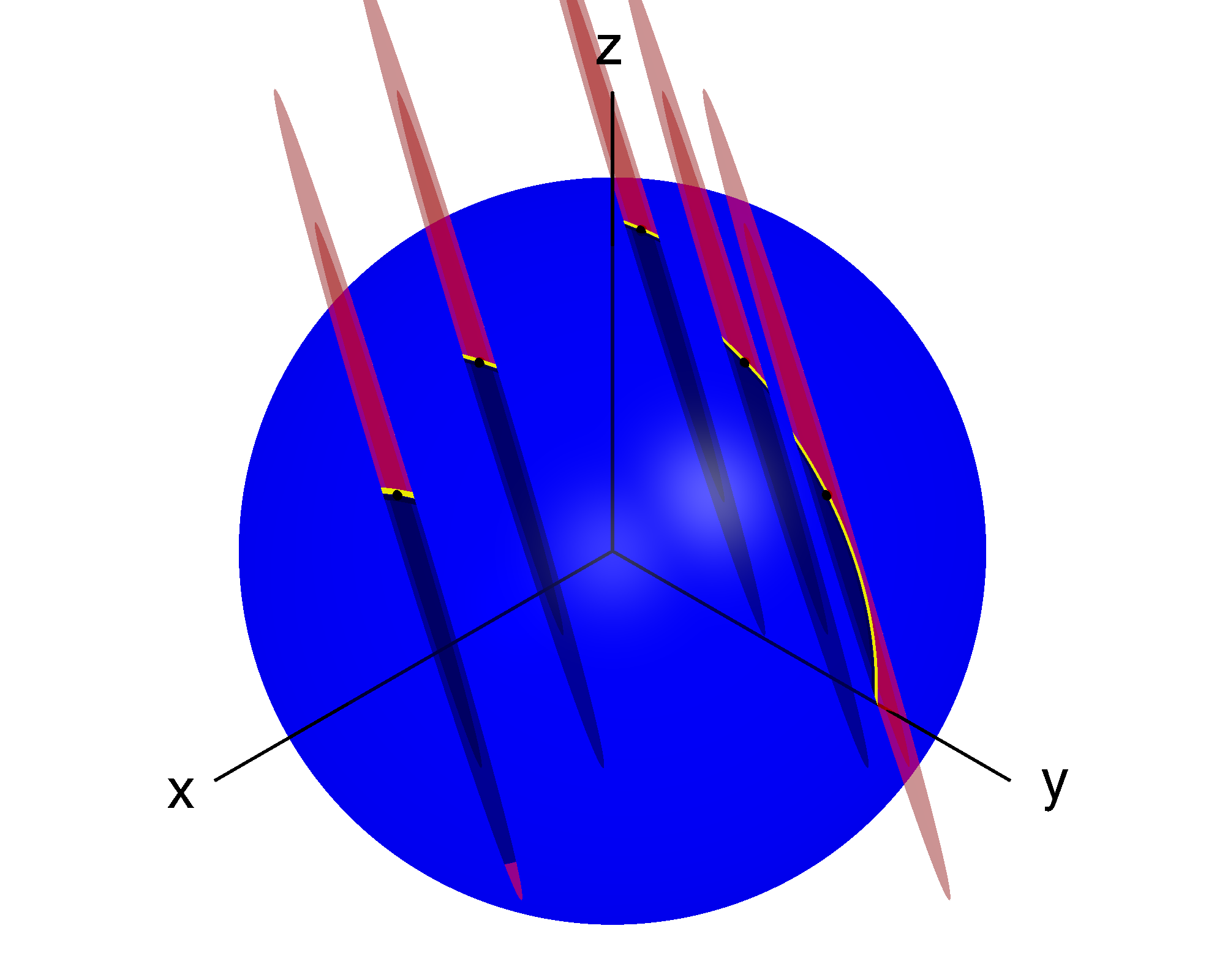}
\caption{\label{fig:ssky_ellipses}
Metric ellipsoids of the supersky metric $\mat g\uss$, at 5 example points, for $T = 4$~days and $\mu\umax = 30$.
Their intersections with the sky sphere $|\vec n| = 1$ reproduce the physical sky metric.
}
\end{figure}

\begin{figure*}
\centering
\includegraphics[width=\linewidth]{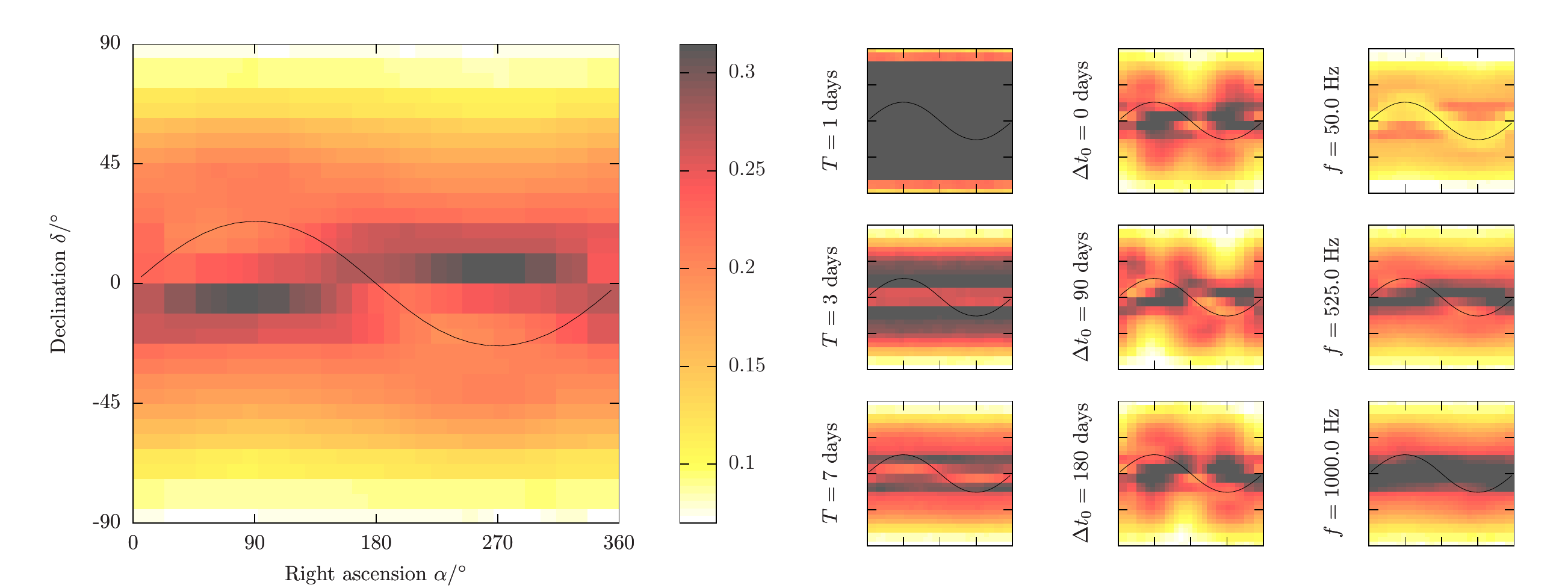}
\caption{\label{fig:mu_sm_f1dot_H1_twoF_ss_mu0p2}
Median magnitude of the relative error $|\relerr{\mu\utwoF}{\mu\uss}|$, as a function of $\alpha$ and $\delta$, for $\mu\utwoF \le 0.2$.
The ecliptic equator is over-plotted in black.
Left: median $|\relerr{\mu\utwoF}{\mu\uss}|$ over time-spans $1 \le T \le 29$~days, and all simulation reference times $\Delta t_0$ and maximum frequencies $f\umax$; see Appendix~\ref{sec:numer-simul}.
Right: median $|\relerr{\mu\utwoF}{\mu\uss}|$ at fixed values of $T$, $\Delta t_0$, and $f\umax$; axis ranges and color values are the same as for the left-hand-side plot.
Only first spindown is used.
}
\end{figure*}

In this paper, to linearize the phase metric $\phi(t, \vec\lambda)$ with respect to the sky coordinates $\vec n$, we simply relax the constraint that $|\vec n| = 1$, and instead consider each of the three components of $\vec n$ to be independent.
It follows that the phase variation $d\phi(t, \vec\lambda)$ is independent of sky position:
\begin{equation}
\label{eq:phi-deriv-nxyz}
\mathrm{d}\phi \propto \vec r(t) \cdot \mathrm{d}\vec n + \dots \,.
\end{equation}
This idea is the foundation of the parameter-space metric described in this paper.
We refer to the phase metric expressed in the three sky coordinates $\vec n\in\mathbb{R}^3$ as the \emph{supersky} metric $\mat g\uss$; it is given e.g. in equatorial coordinates by
\begin{equation}
\label{eq:supersky-metric}
\mat g\uss = \begin{pmatrix}
g_{n\ux,n\ux} & g_{n\ux,n\uy} & g_{n\ux,n\uz} & g_{n\ux,f} & g_{n\ux,\dot f} & \cdots \\
g_{n\ux,n\uy} & g_{n\uy,n\uy} & g_{n\uy,n\uz} & g_{n\uy,f} & g_{n\uy,\dot f} & \cdots \\
g_{n\ux,n\uz} & g_{n\uy,n\uz} & g_{n\uz,n\uz} & g_{n\uz,f} & g_{n\uz,\dot f} & \cdots \\
g_{n\ux,f} & g_{n\uy,f} & g_{n\uz,f} & g_{f,f} & g_{f,\dot f} & \cdots \\
g_{n\ux,\dot f} & g_{n\uy,\dot f} & g_{n\uz,\dot f} & g_{f,\dot f} & g_{\dot f,\dot f} & \cdots \\
\vdots & \vdots & \vdots & \vdots & \vdots & \ddots
\end{pmatrix} \,,
\end{equation}
where the elements $g_{\lambda_i,\lambda_j}$ are given by Eq.~\eqref{eq:phase-metric-def}.

Geometrically, relaxing the constraint $|\vec n| = 1$ amounts to embedding the 2-dimensional physical sky in 3-dimensional space $\mathbb{R}^3$, as illustrated in Figure~\ref{fig:ssky_ellipses}.
The physical sky metric is recovered by reimposing the restriction $|\vec n| = 1$, which is equivalent to finding the intersection of the
sky sphere with a supersky metric ellipsoid centered on a point on the sky sphere.
From Figure~\ref{fig:ssky_ellipses} it is evident that, while the supersky metric ellipsoids have the same shape and orientation, regardless of their location, their intersections with the sky sphere produce shapes of differing sizes and orientations.
This confirms that the metric is flat in the supersky $\vec n \in \mathbb{R}^3$, but not on the physical sky $|\vec n|=1$.

Figures~\ref{fig:mu_re_f1dot_H1_twoF_ss_mu0p2} and~\ref{fig:mu_re_f1dot_H1_twoF_ss_mu0p6} plot the relative error $\relerr{\mu\utwoF}{\mu\uss}$ between mismatch predicted by the supersky metric, and calculated from the $\calF$-statistic, for $\mu\utwoF \le 0.2$ and $0.2 \le \mu\utwoF \le 0.6$ respectively.
For template placement, we are most interested in the median error, and whether the supersky metric significantly under-estimates the $\calF$-statistic mismatch, which would lead to under-covering of parameter-space regions.
For $\mu\utwoF \le 0.2$, the median error is $|\relerr{\mu\utwoF}{\mu\uss}| \lesssim 0.3$ for $T \gtrsim 2$~days, and $|\relerr{\mu\utwoF}{\mu\uss}| \lesssim 0.2$ for $T \gtrsim 7$~days; the 25th--75th percentile range (error bars) is within $\sim 0.5$ for $T \gtrsim 1$~day, and within $\sim 0.2$ for $T \gtrsim 40$~days.
The supersky metric over-estimates the $\calF$-statistic mismatch on average [$\relerr{\mu\utwoF}{\mu\uss} < 0$], leading to slightly conservative template placement.
Only for a small number of trials will the supersky metric significantly under-estimate the $\calF$-statistic mismatch [$\relerr{\mu\utwoF}{\mu\uss} > 0$], e.g.\ at $T \sim 3$~days, $\relerr{\mu\utwoF}{\mu\uss} > 0.5$ for 2.5\% of trials (above dashed line).
For $0.2 \le \mu\utwoF \le 0.6$, the median error magnitude is only slightly worse, $|\relerr{\mu\utwoF}{\mu\uss}| \lesssim 0.25$.

Figure~\ref{fig:mu_sm_f1dot_H1_twoF_ss_mu0p2} plots the median error magnitude $|\relerr{\mu\utwoF}{\mu\uss}|$ as a function of sky position, over the full ranges of simulation parameters, and at fixed values of $T$, $\Delta t_0$, and $f\umax$.
Generally, the $|\relerr{\mu\utwoF}{\mu\uss}|$ are largest in two bands above and below the ecliptic equator (plotted in black), and smallest near the poles.
The area of the sky where the error is large decreases as a function of $T$, increases as a function of $f\umax \lesssim 500$~Hz, and is a more complicated function of the reference time of the data.

While testing the supersky metric by generating an actual template bank is beyond the scope of this paper, we believe the performance of the metric, demonstrated here, is sufficiently accurate for this task.
That the phase metric, despite its approximations, still closely models the $\calF$-statistic mismatch (even at large mismatches of 0.6) demonstrates the relative importance of maintaining phase coherence (as opposed to amplitude consistency) between a gravitational-wave pulsar signal and a search template.
In Section~\ref{sec:comp-alpha-delta}, we demonstrate the suitability of the coordinates $\vec n$ as sky coordinates, as opposed to the angular coordinates $(\alpha, \delta)$.

\subsection{Comparison to the $\calF$-statistic metric}\label{sec:comparison-prix-2007}

\begin{figure}
\centering
\includegraphics[width=\linewidth]{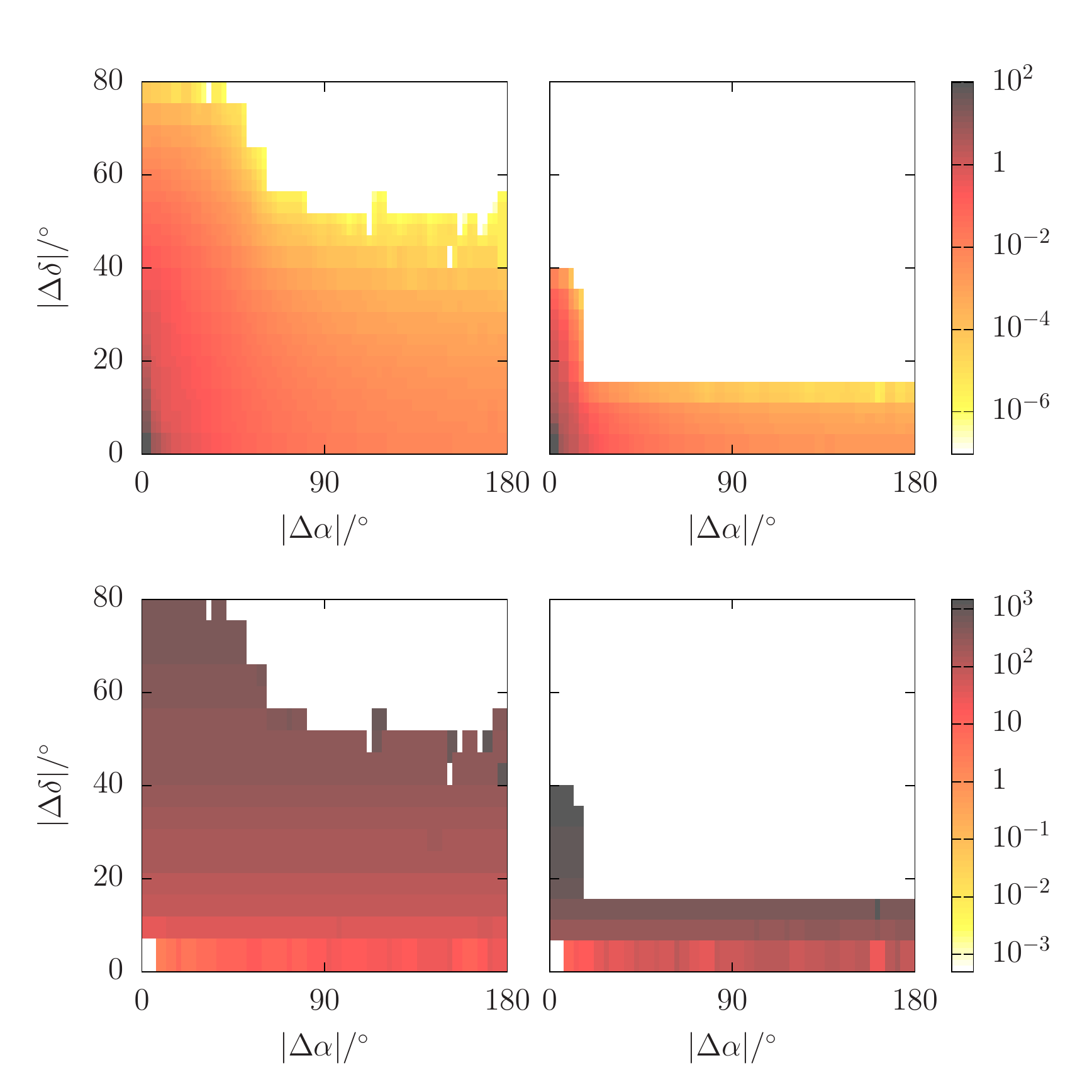}
\caption{\label{fig:offset_maps}
Probability density of simulated coordinate offsets $|\Delta\alpha|$ and $|\Delta\delta|$ (top row), and the minimum of $\Delta\hat\Omega$~[Eq.~\eqref{eq:Delta-Omega-hat-def}] at fixed $|\Delta\alpha|$ and $|\Delta\delta|$ (bottom row), over all simulation reference times and maximum frequencies, and at $T = 1$~day (left column), and $T = 7$~days (right column).
}
\end{figure}

\begin{figure}
\centering
\subfloat[]{\includegraphics[width=0.5\linewidth]{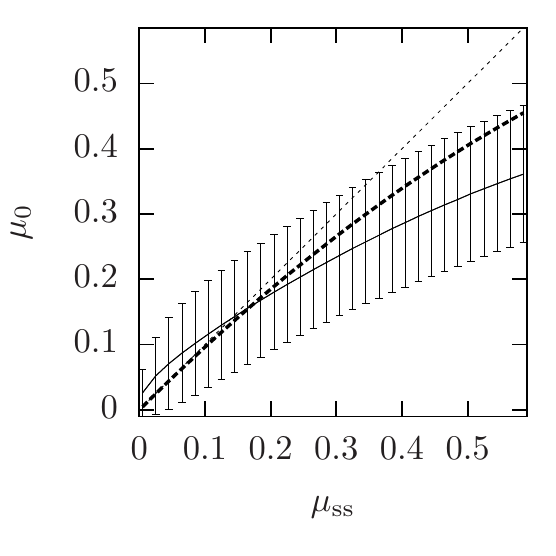}\label{fig:mu_twoF_vs_mu_ss_T1}}
\subfloat[]{\includegraphics[width=0.5\linewidth]{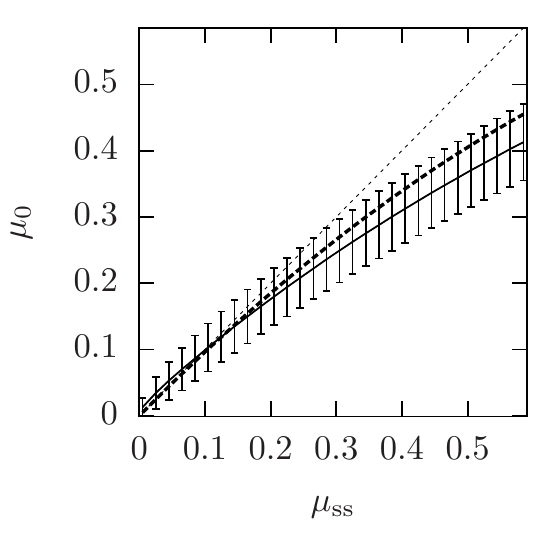}\label{fig:mu_twoF_vs_mu_ss_T7}}
\caption{\label{fig:mu_twoF_vs_mu_ss}
Mean (solid line) and standard deviation (error bars) of $\mu\utwoF$ as a function of $\mu\uss$, over all simulation reference times and maximum frequencies, and for \protect\subref{fig:mu_twoF_vs_mu_ss_T1}~$T = 1$~day, and \protect\subref{fig:mu_twoF_vs_mu_ss_T7}~$T = 7$~days.
The dotted lines plot $\mu\utwoF = \mu\uss$; the dashed lines plot $\mu\utwoF = \mu\uss - 0.38 \mu\uss^{2}$.
}
\end{figure}

The results presented in the previous section may be compared to similar simulations performed in~\cite{Prix.2007a}.
Figure~15 of that paper compares measured $\calF$-statistic mismatches, $\mu\utwoF$, to mismatches predicted by the full $\calF$-statistic metric, $\mu\uFmetr$; while Figures~4 and~5 of that paper compare $\mu\uFmetr$ to mismatches predicted by the phase metric, which is essentially $\mu\uss$.
Taken together, these three figures show, for mismatches up to 0.5, relative errors $\relerr{\mu\utwoF}{\mu\uFmetr} \lesssim 0.05$ and $\relerr{\mu\uFmetr}{\mu\uss} \lesssim 0.4$ for $T \sim 1$~day, and $\relerr{\mu\utwoF}{\mu\uFmetr} \sim 0 \pm 0.01$ and $\relerr{\mu\uFmetr}{\mu\uss} \sim 0 \pm 0.05$ for $T \gtrsim 7$~days.

While the simulations of~\cite{Prix.2007a} suggest better agreement between the $\calF$-statistic and the phase/supersky metric than is shown in e.g.\ Figure~\ref{fig:mu_re_f1dot_H1_twoF_ss_mu0p2}, one must first account for a subtle difference between the two sets of simulations: the distributions of the randomly-generated sky coordinate offsets $|\Delta\alpha|$ and $|\Delta\delta|$ from which the mismatches are calculated.
The simulations in~\cite{Prix.2007a} drew offsets uniformly distributed in $|\Delta\alpha|$ and $|\Delta\delta|$, and furthermore excluded ``large'' sky offsets, quantified by
\begin{equation}
\label{eq:Delta-Omega-hat-def}
\Delta\hat\Omega \approx 10^{-4} f T \sqrt{(\Delta\alpha \cos\delta)^2 + (\Delta\delta)^2} \gtrsim 5 \,.
\end{equation}
In contrast, the simulations in this paper sample offsets uniformly in coordinates defined by the eigenvectors of the supersky metric, i.e.\ the axes of the metric ellipsoids (see Appendix~\ref{sec:numer-simul} for details).
As seen in Figure~\ref{fig:ssky_ellipses}, the supersky metric ellipsoids are typically highly elongated along one semi-major axis.
Giving equal weight to each metric semi-major axis when sampling offsets, as is done in this paper, leads to a greater number of large sky offsets than is achieved by uniform sampling in $|\Delta\alpha|$ and $|\Delta\delta|$ (see discussion in Section IV\,E\,2 of~\cite{Prix.2007a}).
Figure~\ref{fig:offset_maps} plots, at $T = 1$ and 7 days, the probability density of $|\Delta\alpha|$ and $|\Delta\delta|$, and the minimum sampled value of $\Delta\hat\Omega$, for simulations presented in this paper.
The distributions of $|\Delta\alpha|$ and $|\Delta\delta|$ are far from uniform, and contain few points that would satisfy the $\Delta\hat\Omega \lesssim 5$ cut-off used in~\cite{Prix.2007a}.
We expect differences between the $\calF$-statistic and the supersky metric to be magnified at larger coordinate offsets, and so it is unsurprising that the errors shown in e.g.\ Figure~\ref{fig:mu_re_f1dot_H1_twoF_ss_mu0p2} are larger than those found in~\cite{Prix.2007a}.

Figure~\ref{fig:mu_twoF_vs_mu_ss} plots the mean and standard deviation of $\mu\utwoF$ versus $\mu\uss$, at $T = 1$ and 7 days.
It is qualitatively similar to Figure~10 of~\cite{Prix.2007a}, which plots $\mu\utwoF$ versus $\mu\uFmetr$ at $T = 0.5$ and 2.5 days.
In both cases, higher-order terms in $\Delta\vec\lambda$, neglected in the derivation of Eq.~\eqref{eq:mismatch-metr}, result in both $\mu\uFmetr$ and $\mu\uss$ over-estimating $\mu\utwoF$ at larger mismatches.
The empirical fit $\mu\utwoF = \mu\uFmetr - 0.38 \mu\uFmetr^{2}$ to the behavior of the full $\calF$-statistic metric, found in~\cite{Prix.2007a}, is not as good a fit to the behavior of $\mu\uss$, suggesting that the approximations made in deriving the phase metric lead to further over-estimation of $\mu\utwoF$.

\subsection{Comparison to sky metric in $\alpha$--$\delta$ coordinates}\label{sec:comp-alpha-delta}

\begin{figure}
\centering
\includegraphics[width=\linewidth]{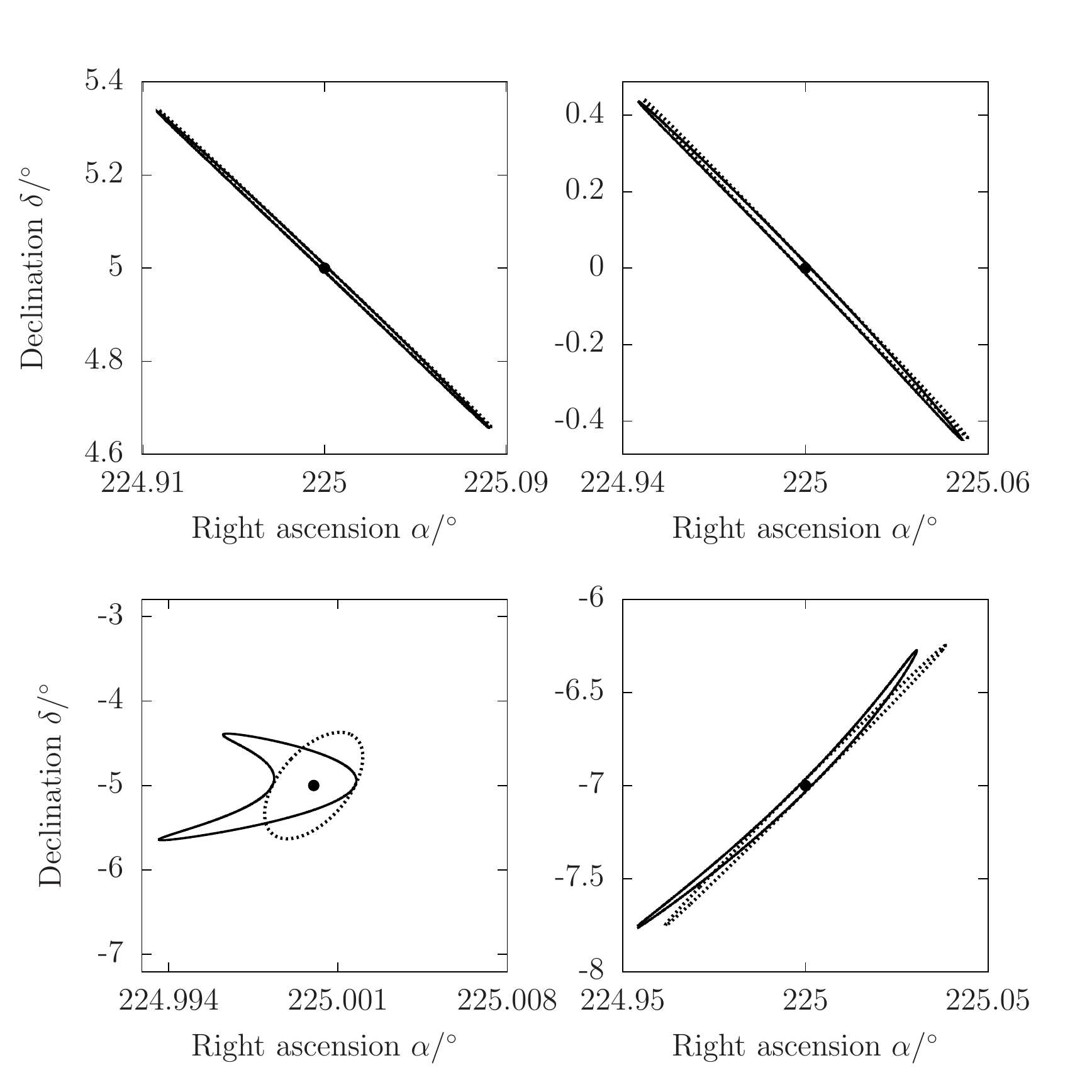}
\caption{\label{fig:alpha_delta_ellipses}
True mismatch region $\mu \le \mu\umax$ of the sky metric (solid line), and the local metric ellipse in $(\alpha, \delta)$ given by Eq.~\eqref{eq:mismatch-sky-def} (dotted line), at 4 example points (black dots), for $T = 2$~days and $\mu\umax = 0.3$.
}
\end{figure}

\begin{figure}
\centering
\subfloat[]{\includegraphics[width=\linewidth]{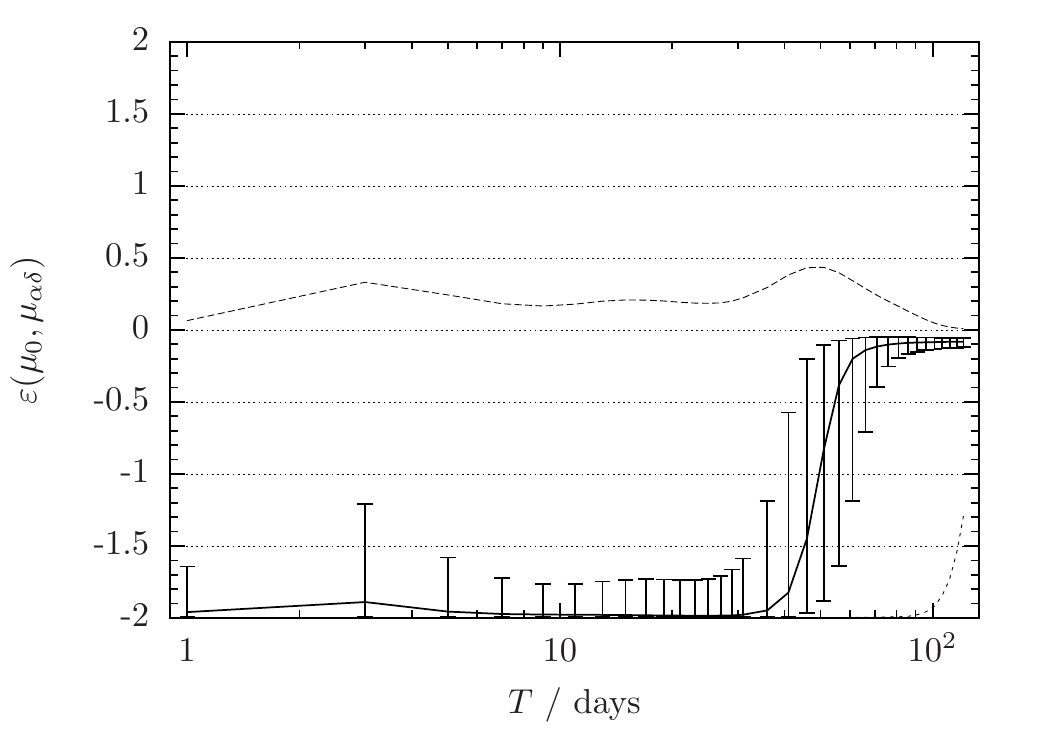}\label{fig:mu_re_f1dot_H1_twoF_ssad_mu0p2}}\\
\subfloat[]{\includegraphics[width=\linewidth]{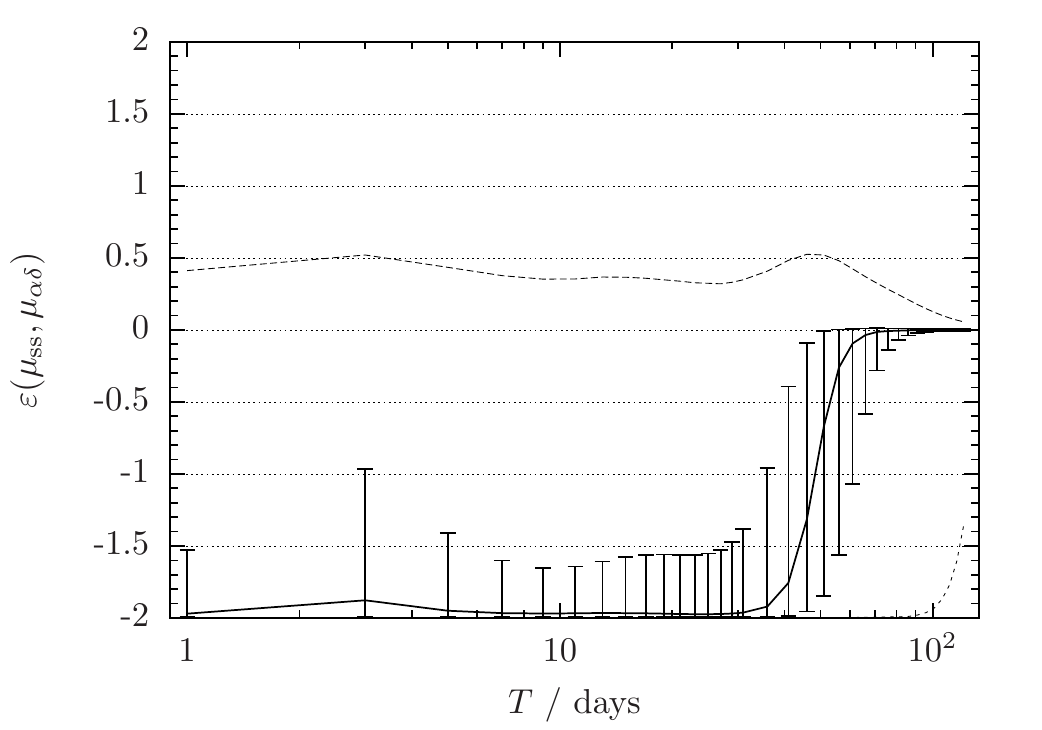}\label{fig:mu_re_f1dot_H1_ss_ssad_mu0p2}}
\caption{\label{fig:mu_re_f1dot_H1_ssad_mu0p2}
Relative errors \protect\subref{fig:mu_re_f1dot_H1_twoF_ssad_mu0p2}~$\relerr{\mu\utwoF}{\mu\ussad}$ and \protect\subref{fig:mu_re_f1dot_H1_ss_ssad_mu0p2}~$\relerr{\mu\uss}{\mu\ussad}$ between mismatches predicted by the sky metric in $\alpha$--$\delta$ coordinates, $\mu\ussad$, predicted by the supersky metric, $\mu\uss$, and calculated from the $\calF$-statistic, $\mu\utwoF$, as a function of $T$, for $\mu\utwoF \le 0.2$ and $\mu\uss \le 0.2$ respectively.
Plotted are the median (solid line), the 25th--75th percentile range (error bars), and the 2.5th (dotted line) and 97.5th (dashed line) percentiles.
Only first spindown is used.
}
\end{figure}

In contrast to the supersky coordinates $\vec n$, the angular coordinates $(\alpha, \delta)$ are a poor choice of sky coordinates for the purpose of predicting mismatch.
This is because, since the sky metric expressed in $\alpha$--$\delta$ coordinates is itself a function of $\alpha$ and $\delta$, the mismatch in these coordinates is generally calculated by evaluating the metric at a given point, e.g.\ $(\alpha_0, \delta_0)$, and computing
\begin{equation}
\label{eq:mismatch-sky-def}
\mu\ussad \approx
\begin{pmatrix} \alpha - \alpha_0 \\ \delta - \delta_0 \\ \Delta f \\ \vdots \end{pmatrix}
\cdot \mat g |_{\alpha=\alpha_0, \delta=\delta_0}
\begin{pmatrix} \alpha - \alpha_0 \\ \delta - \delta_0 \\ \Delta f \\ \vdots \end{pmatrix} \,.
\end{equation}
As can be seen in Figure~\ref{fig:ssky_ellipses}, this is often a very poor approximation, since the sky metric can change noticeably as a function of $\alpha$ and $\delta$.
To illustrate, Figure~\ref{fig:alpha_delta_ellipses} plots, for 4 example points, the true mismatch region of the sky metric, given by the intersection of supersky metric ellipsoid and sky sphere, alongside the local metric ellipse, which is found from the intersection of the supersky metric ellipsoid with a plane tangent to the sky sphere at the chosen point.
The agreement between the two changes from very good (top-left plot) to very poor (bottom-left plot) within a 10-degree change in $\delta$.

Figure~\ref{fig:mu_re_f1dot_H1_ssad_mu0p2} plots the relative errors $\relerr{\mu\utwoF}{\mu\ussad}$ and $\relerr{\mu\uss}{\mu\ussad}$ between mismatches predicted by the sky metric in $\alpha$--$\delta$ coordinates, by the supersky metric, and calculated from the $\calF$-statistic.
While the sky metric in $\alpha$--$\delta$ coordinates begins to perform better at longer $T \gtrsim 60$~days, it performs extremely poorly at shorter $T$.
This is a consequence of the poor approximation in Eq.~\eqref{eq:mismatch-sky-def} which, as shown in Figure~\ref{fig:alpha_delta_ellipses}, can cause the sky region covered by the local metric ellipsoid at $(\alpha_0, \delta_0)$ to be very different from the true mismatch region.
This difference in covered sky regions tends to show up as an over-estimation of e.g.\ $\mu\utwoF$ by $\mu\ussad$ [$\relerr{\mu\utwoF}{\mu\ussad} < 0$], as illustrated in Figure~\ref{fig:relerr_cartoons}.

\section{The reduced supersky metric}\label{sec:reduced-supersky}

The fundamental idea behind the supersky metric, presented in Section~\ref{sec:supersky-metric}, is also its main drawback when it comes to template placement: it embeds the 2-dimensional physical sky in 3-dimensional space.
This means that we cannot simply fill the 3-dimensional supersky space with templates, since only a small fraction of them will correspond to physical sky positions (i.e.\ satisfy $|\vec n| = 1$).
Instead, we must find a way to reduce the dimensionality of the supersky metric to 2 dimensions, while preserving flatness, and without making assumptions that introduce significant errors between the supersky and $\calF$-statistic mismatches.

\begin{figure}
\includegraphics[width=\linewidth]{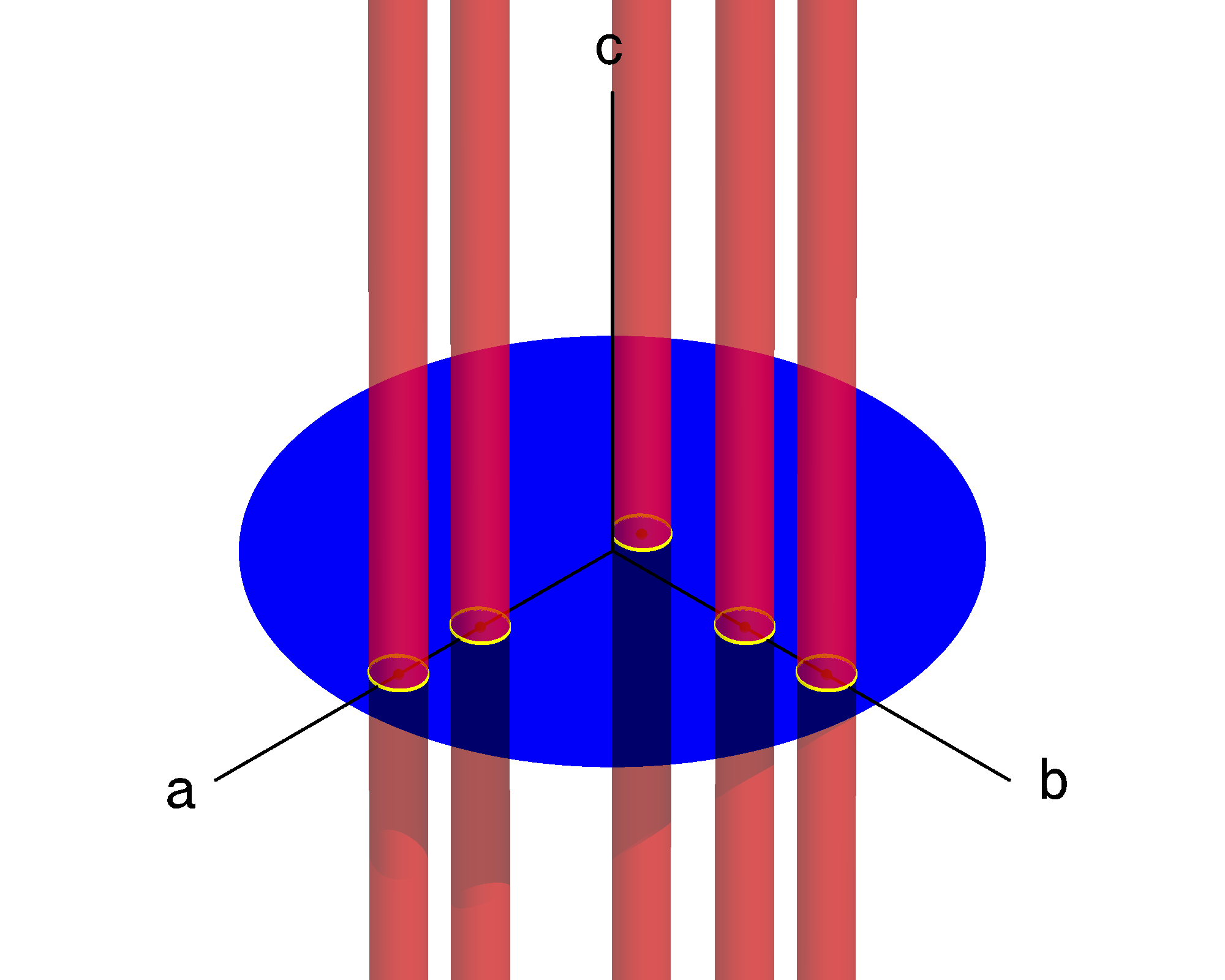}
\caption{\label{fig:aligned_ssky_ellipses}
Metric ellipsoids of the supersky metric, in aligned coordinates $(n\ua, n\ub, n\uc)$ [see Section~\ref{sec:diagonalizing-metric}], for $T = 4$~days and $\mu\umax = 5$, at 5 example points.
By removing the $c$ dimension, we produce a projected metric on the 2-dimensional $a$--$b$ plane.
}
\end{figure}

The approach taken in this paper is to derive a new set of sky coordinates $(n\ua, n\ub, n\uc)$, and frequency and spindown coordinates $(\nu, \dot \nu, \dots)$, such that the supersky metric in these coordinates, $\mat g\uss\dg$, is (nearly) diagonal:
\begin{equation}
\label{eq:g-ss-diagonal}
\mat g\uss\dg = \begin{pmatrix}
g\dg_{n\ua, n\ua} & 0 & 0 & 0 & 0 & \cdots \\
0 & g\dg_{n\ub, n\ub} & 0 & 0 & 0 & \cdots \\
0 & 0 & g\dg_{n\uc, n\uc} & 0 & 0 & \cdots \\
0 & 0 & 0 & g\dg_{\nu, \nu} & g\dg_{\nu, \dot \nu} & \cdots \\
0 & 0 & 0 & g\dg_{\nu, \dot \nu} & g\dg_{\dot \nu, \dot \nu} & \cdots \\
\vdots & \vdots & \vdots & \vdots & \vdots & \ddots \\
\end{pmatrix} \,.
\end{equation}
Suppose that the new sky coordinates are chosen such that the inequalities $g\dg_{n\ua, n\ua} \ge g\dg_{n\ub, n\ub} \ge g\dg_{n\uc, n\uc}$ always hold.
The metric ellipsoids of the supersky metric in these coordinates, shown in Figure~\ref{fig:aligned_ssky_ellipses}, are longest along the $c$ axis; equivalently, the mismatch is most insensitive to changes in the coordinates $n\uc$.
Dropping the $c$ dimension, therefore, will introduce the smallest possible error in calculating mismatches, relative to using the full 3-dimensional supersky metric.
This is geometrically equivalent to projecting the supersky metric onto the 2-dimensional $a$--$b$ plane, as seen Figure~\ref{fig:aligned_ssky_ellipses}.
The \emph{reduced supersky} metric $\mat g\urss$ in the coordinates $(n\ua, n\ub, \nu, \dot \nu, \dots)$ reduces the sky dimensionality to 2 dimensions, while remaining constant.
Its derivation is presented in the remainder of this section.

\subsection{Diagonalization and condition numbers}\label{sec:diag-cond-numb}

\begin{figure}
\centering
\includegraphics[width=\linewidth]{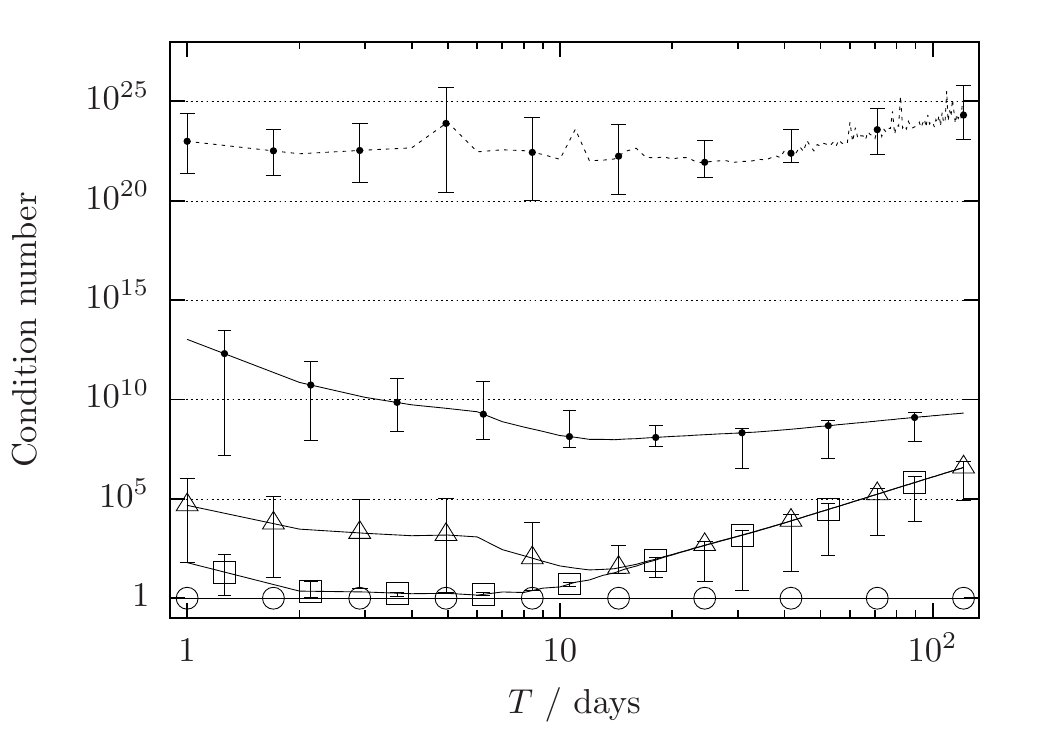}
\caption{\label{fig:condition_nums}
Condition numbers of: the supersky metric, in S.I.\ units, $\mat g\uss$ (points, dotted line), and after diagonal rescaling, $\matbar g\uss$ (points, solid line); the diagonal-rescaled metric $\matbar g\uss\pr$, derived in Section~\ref{sec:modell-orbit-moti} (triangles); and the diagonal-rescaled metrics $\matbar g\uss\ddg$ (squares) and $\matbar g\uss\dg$ (circles), derived in Section~\ref{sec:diagonalizing-metric}.
Plotted as functions of $T$ are the mean (lines) and the minimum-to-maximum range (error bars) of condition numbers over all simulation reference times $\Delta t_0$; see Appendix~\ref{sec:numer-simul}.
Only first spindown is used.
}
\end{figure}

\begin{figure}
\centering
\includegraphics[width=\linewidth]{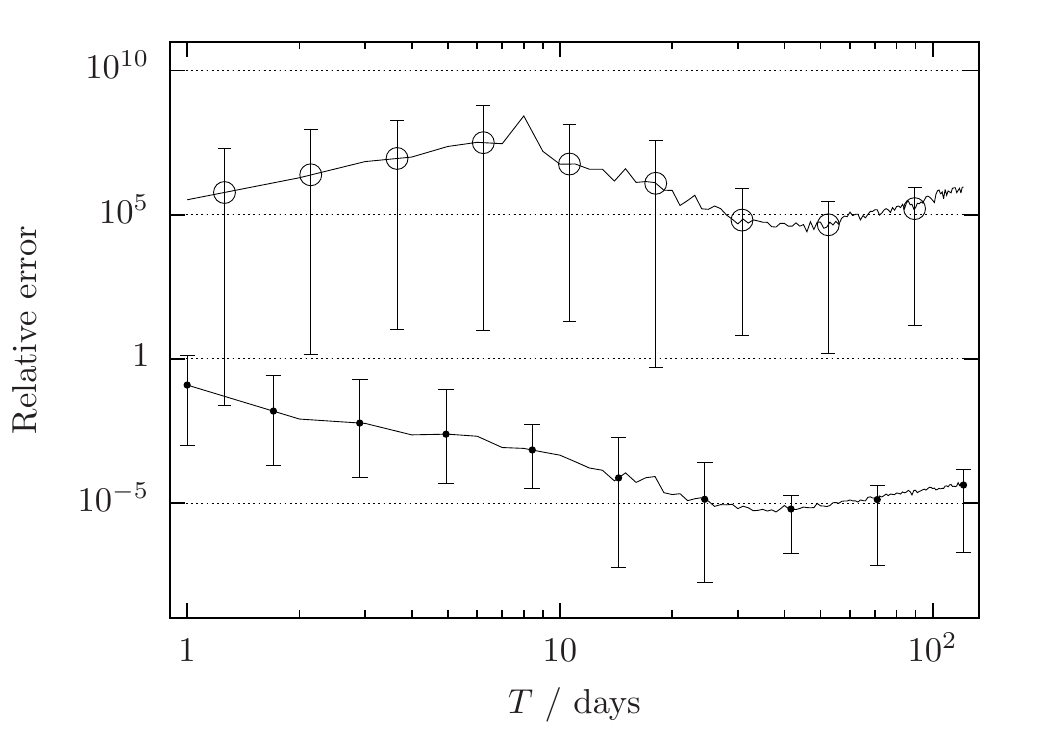}
\caption{\label{fig:smallest_eigval_err}
Relative error in computing the smallest eigenvalue of the supersky metric in S.I.\ units, including first spindown (points), and second spindown (circles).
Plotted as functions of $T$ are the mean (lines) and the minimum-to-maximum range (error bars) of errors over all simulation reference times $\Delta t_0$; see Appendix~\ref{sec:numer-simul}.
}
\end{figure}

A straightforward approach to diagonalizing the supersky metric would be decompose it as $\mat g\uss = \mat Q \mat \Lambda \mat Q\trsp$, where $\mat \Lambda$ is a diagonal matrix of eigenvalues, and $\mat Q$ is an orthogonal matrix whose columns are the corresponding eigenvectors; its transpose $\mat Q\trsp$ gives the linear transform from the original to the new supersky coordinates.
This approach is complicated, however, by a practical difficulty: the numerical stability of the matrix $\mat g\uss$.
A useful measure of numerical stability of a matrix is its \emph{condition number}, which for a real symmetric matrix $\mat g\uss$ is given by the absolute ratio of the largest to the smallest of its eigenvalues~\citep[e.g.][]{Higham.2002}.
Generally, when the condition number is of the same order as the numerical precision of the matrix components, computations using the matrix become unreliable~\cite{Prix.2007a}.
When computed in S.I.\ units (i.e.\ $f$ in Hz, $\dot f$ in Hz/s, etc.), the condition number of $\mat g\uss$, plotted in Figure~\ref{fig:condition_nums}, is $\gtrsim10^{20}$, much larger than the $\sim2\text{ in }10^{16}$ numerical precision of double-precision floating-point computer arithmetic.
Figure~\ref{fig:smallest_eigval_err} shows the effect of the metric's ill-conditionedness, by plotting the relative error $|\lambda\uss - \lambda\uss\dg|/\lambda\uss\dg$ between the smallest eigenvalue $\lambda\uss$ computed from the untransformed metric $\mat g\uss$, and the smallest eigenvalue $\lambda\uss\dg$ computed from the diagonalized metric $\mat g\uss\dg$~[Eq.~\eqref{eq:g-ss-diagonal}] obtained in Section~\ref{sec:reduc-sky-dimens}.
When the metric includes only first spindown, relative errors of $\gtrsim 1$ are possible for $T \lesssim 2$~days; when second spindown is included, the errors increase by several orders of magnitude.

One simple method of reducing the condition number of a matrix $\mat g$ is the following diagonal rescaling:
\begin{equation}
\label{eq:diag-rescaling}
\bar g(\lambda_i, \lambda_j) = \frac{ g(\lambda_i, \lambda_j) }{ \sqrt{ g(\lambda_i, \lambda_i) g(\lambda_j, \lambda_j) } } \,,
\end{equation}
where the $\bar g(\lambda_i, \lambda_j)$ are elements of a rescaled matrix $\matbar g$.
This particular diagonal rescaling reduces the condition number to within a factor $n$ of the smallest condition number achievable by any diagonal rescaling, $n$ being the dimensionality of the matrix~\cite{Higham.2002}.
Even after applying Eq.~\eqref{eq:diag-rescaling} to the supersky metric, resulting in $\matbar g\uss$, its condition number is still of the order $\sim10^{8}$--$10^{15}$ (Figure~\ref{fig:condition_nums}).

We therefore apply a series of transformations to the supersky metric, described in the following sections, which are designed to both diagonalize the metric and further reduce its condition number.

\subsection{Modeling the orbital motion}\label{sec:modell-orbit-moti}

The first transformation takes advantage of a well-known property of the gravitational-wave pulsar signal model: that for coherent time-spans $T$ much shorter than 1~year, the change in signal phase due to the Earth's orbital motion closely resembles a change in phase due to the frequency evolution of the pulsar.
This is because, for $\Delta t \ll T\uo = 1$~year, and assuming a circular orbit about the Sun, the orbital motion of the Earth is modeled by terms involving $\sin 2\pi \Delta t/T\uo$ and  $\cos 2\pi \Delta t/T\uo$; for small $\Delta t/T\uo$, these can be Taylor-expanded as a series of terms in $(\Delta t/T\uo)^s$, which are then absorbed into the frequency and spindown terms in the signal phase [Eq.~\eqref{eq:phase-det}].
This property of the signal model is the basis of the global correlation method~\cite{Pletsch.Allen.2009a,Pletsch.2010a}.

Here, we hypothesize that the similarity between the orbital motion and frequency evolution terms in the signal model leads to a linear relation between the corresponding components in the supersky metric, and hence to an ill-conditioned matrix $\mat g\uss$.
We introduce an intermediate set of coordinates $(\vec n, \ndot f{}\pr)$ which take advantage of this relation, and result in a metric $\mat g\uss\pr$ with a greatly reduced condition number.
The derivation of these coordinates consists of 3 steps: splitting the diurnal and orbital motion of the detector into separate sky coordinates (Section~\ref{sec:split-sider-orbit}), performing a least-squares fit to the orbital motion using the frequency and spindown coordinates (Section~\ref{sec:least-squares-linear}), and recombining the diurnal and orbital motion to recover the supersky coordinates (Section~\ref{sec:recomb-sider-orbit}).

\subsubsection{Splitting the diurnal and orbital motions}\label{sec:split-sider-orbit}

Recall that the supersky coordinates $\vec n$ enter the signal phase $\phi(t, \vec\lambda)$ through the expression $\vec r(t) \cdot \vec n$~[Eq.~\eqref{eq:phase-det}].
We split $\vec r(t) = \vec r\us(t) + \vec r\uo(t)$ into its diurnal and orbital components, $\vec r\us(t)$ and $\vec r\uo(t)$ respectively:
\begin{equation}
\label{eq:split-sid-orb-1}
\vec r(t) \cdot \vec n = \vec r\us(t) \cdot \vec n + \vec r\uo(t) \cdot \vec n \,.
\end{equation}
We now relax the constraint that the $\vec n$ which multiplies $r\us(t)$ is the same $\vec n$ which multiplies $r\uo(t)$.
Instead, we introduce new sky position vectors, $\vec n\us$ and $\vec n\uo$, and write
\begin{equation}
\label{eq:split-sid-orb-2}
\vec r(t) \cdot \vec n = \vec r\us(t) \cdot \vec n\us + \vec r\uo(t) \cdot \vec n\uo \,,
\end{equation}
where $\vec n\us$ and $\vec n\uo$ are now treated as independent sets of coordinates.
If we now write $\vec r\us(t) = [r\usx(t), r\usy(t), r\usz]$ and $\vec n\us = (n\usx, n\usy, n\usz)$ in equatorial coordinates, and
$\vec r\uo(t) = [r\uoX(t), r\uoY(t), r\uoZ(t)]$ and $\vec n\uo = (n\uoX, n\uoY, n\uoZ)$ in ecliptic coordinates, we have
\begin{equation}
\begin{split}
\label{eq:split-sid-orb-3}
\vec r(t) \cdot \vec n &= r\usx(t) n\usx + r\usy(t) n\usy + r\usz n\usz \\
&\quad + r\uoX(t) n\uoX + r\uoY(t) n\uoY + r\uoZ(t) n\uoZ \,.
\end{split}
\end{equation}
Since $r\usz$ is a constant (i.e.\ there is no motion of the detector with respect to the Earth in the equatorial $z$ direction), $g(n\usz, \lambda_i) = 0$ for any coordinate $\lambda_i$, and so we ignore the term $r\usz n\usz$.
(Since, however, the Earth's orbit does include motion in the ecliptic $Z$ direction, due to e.g.\ its interaction with the Moon, we do not neglect the term $r\uoZ(t) n\usZ$.)
We are left with 5 independent sky coordinates,
\begin{equation}
\label{eq:vec-n-ess}
\vec n\uess = (n\usx, n\usy, n\uoX, n\uoY, n\uoZ) \,,
\end{equation}
and we write the metric in these \emph{expanded supersky} coordinates as $\mat g\uess$.
Just as the supersky coordinates $\vec n$ embedded the 2-dimensional physical sky in 3 dimensions, the expanded supersky coordinates embed the 3-dimensional supersky in 5 dimensions.
Re-imposing the constraint $\vec n\us = \vec n\uo$ recovers the supersky metric $\mat g\uss$.

\subsubsection{Least-squares linear fit to the orbital motion}\label{sec:least-squares-linear}

We apply the diagonal rescaling of Eq.~\eqref{eq:diag-rescaling} to the expanded supersky metric:
\begin{equation}
\label{eq:bar-g-ss-ess}
\bar g\uess(\lambda_i, \lambda_j) = \frac{ g\uess(\lambda_i,\lambda_j) }{ \sqrt{ g\uess(\lambda_i,\lambda_i) g\uess(\lambda_j,\lambda_j) } } \,,
\end{equation}
where $\lambda_i, \lambda_j \in \vec\lambda = ( \vec n\uess, \ndot f )$.
We now hypothesize an (approximately) linear relationship between the rows of $\matbar g\uess$ corresponding to orbital motion in the ecliptic $X$ and $Y$ directions, and the rows of $\matbar g\uess$ corresponding to frequency evolution, i.e.\ we write
\begin{multline}
\label{eq:bar-g-ess-fit}
\bar g\uess(\lambda_i, n\uoSigma) = \sum_{s=0}^{s\umax} \bar g\uess(\lambda_i, \ndot f) \, C(\ndot f, n\uoSigma) \\ + \delta\bar g\uess(\lambda_i, n\uoSigma) \,,
\end{multline}
where $\sigma \in \{X,\,Y\}$, $C(\ndot f, n\uoSigma)$ are the components of a $(1 + s\umax) \times 2$ matrix of fitting coefficients $\mat C$, and $\delta\bar g\uess(\lambda_i, n\uoSigma)$ are components of a $(6 + s\umax) \times 2$ matrix of residuals $\bm{\delta}\matbar
g\uess$. Equations~\eqref{eq:bar-g-ess-fit} are overdetermined, since they represent $(6 + s\umax) \times 2$ equations, one for each element $\bar g\uess(\lambda_i, n\uoSigma)$, for only $(1 + s\umax) \times 2$ unknowns $C(\ndot f, n\uoSigma)$.
Using linear least squares for each $\sigma\in\{X,Y\}$, i.e.\ minimizing the objective function $\sum_i \left[\delta\bar g\uess(\lambda_i, n\uoSigma)\right]^2$ over the vector $C(\ndot f,n\uoSigma)$ at fixed $\sigma$, yields the fitting coefficients
\begin{equation}
\label{eq:bar-g-ess-fit-C}
C(\ndot f, n\uoSigma) = \sum_{s'=0}^{s\umax} A^{-1}(\ndot f, \ndot[s'] f) \, B(\ndot[s'] f, n\uoSigma) \,,
\end{equation}
where the $(1 + s\umax) \times 2$ matrix $\mat B$ has components
\begin{equation}
\label{eq:bar-g-ess-fit-B}
B(\ndot f, n\uoSigma) = \sum_{i} \bar g\uess(\ndot f, \lambda_i)\, \bar g\uess(\lambda_i, n\uoSigma)\,,
\end{equation}
and $A^{-1}(\ndot f, \ndot[s']f)$ are the elements of the inverse of the symmetric $(1 + s\umax) \times (1 + s\umax)$ matrix $\mat A$, with components
\begin{equation}
\label{eq:bar-g-ess-fit-A}
A(\ndot f, \ndot[s']f) = \sum_{i} \bar g\uess(\ndot f, \lambda_i) \, \bar g\uess(\lambda_i, \ndot[s']f) \,.
\end{equation}

Next, we perform a coordinate transformation which results in a metric $\matbar g\uess\ppr$.
In this metric, the linear fits to $\bar g\uess(\lambda_i, n\uoSigma)$ are subtracted, leaving only the residuals:
\begin{equation}
\label{eqs:bar-g-ess-ppr}
\bar g\uess\ppr(\lambda_i, n\uoSigma) = \delta\bar g\uess(\lambda_i, n\uoSigma) \,.
\end{equation}
This is achieved by the coordinate transform
\begin{multline}
\label{eq:ndot-f-ppr-def}
\ndot f{}\ppr = \ndot f + \sum_{\sigma\in\{X,Y\}} n\uoSigma \, C(\ndot f, n\uoSigma) \\ \times \sqrt{ \frac{ g\uess(n\uoSigma, n\uoSigma) }{ g\uess(\ndot f,\ndot f) } }\,,
\end{multline}
with the sky coordinates $\vec n\uess$ remaining unchanged.
The new frequency and spindown coordinates $\ndot f{}\ppr$ are now linear functions of sky position, similar to the global correlation coordinates [c.f.\ Eqs.~\eqref{eqs:gct-nu}].

\subsubsection{Recombining the diurnal and orbital motions}\label{sec:recomb-sider-orbit}

\begin{figure}
\includegraphics[width=\linewidth]{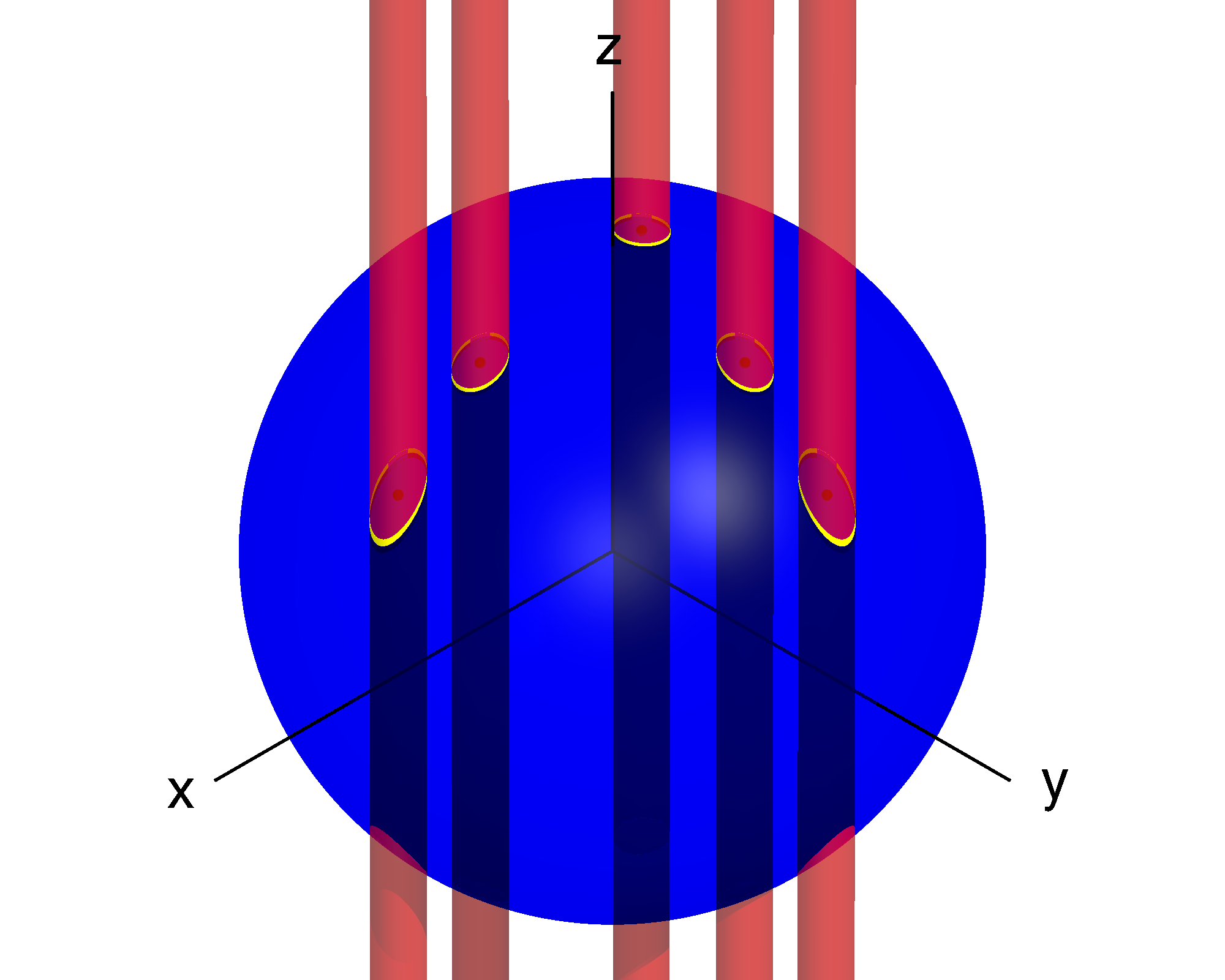}
\caption{\label{fig:residual_ssky_ellipses}
Metric ellipsoids of the transformed supersky metric $\mat g\uss\pr$, in equatorial coordinates $\vec n$, for $T = 4$~days and $\mu\umax = 5$, at 5 example points, and their intersections with the sky sphere $|\vec n| = 1$.
}
\end{figure}

Finally, we re-impose the constraint $\vec n\us = \vec n\uo$, and reverse the diagonal rescaling given by Eq.~\eqref{eq:bar-g-ss-ess}.
This gives the final product of this section, the metric $\mat g\uss\pr$.
The coordinates of this metric are $(\vec n, \ndot f{}\pr)$, where $\vec n$ is the 3-dimensional sky position vector, and
\begin{equation}
\label{eq:ndotbar-f-pr-def}
\ndot f{}\pr = \ndot f + \vec\Gamma^s \cdot \vec n \,.
\end{equation}
The components of the vectors $\vec\Gamma^s$ are given in ecliptic coordinates by
\begin{subequations}
\label{eqs:Gamma-s-ecl-XYZ}
\begin{align}
\Gamma^s\uX &= C(\ndot f, n\uoX) \sqrt{ \frac{ g\uess(n\uoX,n\uoX) }{ g\uess(\ndot f,\ndot f) } } \,, \\
\Gamma^s\uY &= C(\ndot f, n\uoY) \sqrt{ \frac{ g\uess(n\uoY,n\uoY) }{ g\uess(\ndot f,\ndot f) } } \,, \\
\Gamma^s\uZ &= 0 \,.
\end{align}
\end{subequations}
and in equatorial coordinates by
\begin{subequations}
\label{eqs:Gamma-s-equ-xyz}
\begin{align}
\Gamma^s\ux &= \Gamma^s\uX\,,\\
\Gamma^s\uy &= \Gamma^s\uY \, \cos\epsilon \,,\\
\Gamma^s\uz &= \Gamma^s\uY \, \sin\epsilon \,,
\end{align}
\end{subequations}
where $\epsilon$ is the Earth's inclination angle with respect to the ecliptic $Z$-direction.

In summary, the coordinate transformation presented in this section consists of the following steps:
\begin{enumerate}[(i)]

\item Compute the phase metric $\mat g\uess$, using Eq.~\eqref{eq:phase-metric-def}, in the expanded supersky coordinates of Eq.~\eqref{eq:vec-n-ess}.
The signal phase is found by substituting Eq.~\eqref{eq:split-sid-orb-3} in Eq.~\eqref{eq:phase-det}.

\item Rescale the metric using Eq.~\eqref{eq:bar-g-ss-ess}, which gives the metric $\matbar g\uess$.

\item Compute the matrices $\mat A$, using Eq.~\eqref{eq:bar-g-ess-fit-A}, and $\mat B$, using Eq.~\eqref{eq:bar-g-ess-fit-B}.

\item Compute the matrix $\mat C$ from Eq.~\eqref{eq:bar-g-ess-fit-C}.

\item Compute the vectors $\vec \Gamma^s$ from Eqs.~\eqref{eqs:Gamma-s-ecl-XYZ} and \eqref{eqs:Gamma-s-equ-xyz}.

\item Apply the inverse of the coordinate transformation given by Eq.~\eqref{eq:ndotbar-f-pr-def} to the supersky metric $\mat g\uss$, which gives the metric $\mat g\uss\pr$.

\end{enumerate}
The condition number of this metric after diagonal rescaling, $\matbar g\uss\pr$, is plotted in Figure~\ref{fig:condition_nums}; it is reduced to $\sim 1$--$10^{6}$, and is $\sim 3$--8 orders of magnitude less than the condition number of the untransformed metric, $\matbar g\uss$.

Figure~\ref{fig:residual_ssky_ellipses} plots the metric ellipsoids of $\mat g\uss\pr$, with $\mu\umax = 5$.
The areas of intersections of the ellipsoids of $\mat g\uss\pr$ with the sky sphere $|\vec n| = 1$ are larger (and more circular) than for the ellipsoids of $\mat g\uss$, plotted in Figure~\ref{fig:ssky_ellipses} with $\mu\umax = 30$, which implies a coarser sky metric induced by $\mat g\uss\pr$ than by the untransformed metric, $\mat g\uss$.
This can be intuitively understood as follows: given the constraint $\mu \le \mu\umax$, the resolution in the supersky coordinates, e.g.\ $|\Delta n\ux|$, are inversely proportional to the corresponding supersky metric elements, e.g.\ $g\uss(n\ux, n\ux) (\Delta n\ux)^2 = \mu \le \mu\umax$ implies $|\Delta n\ux| \propto g\uss(n\ux, n\ux)^{-1/2}$.
The magnitude of $g\uss(n\ux, n\ux)$ is proportional to the magnitude of the detector position vector, $|\vec r(t)|^2$, which is dominated by the magnitude of its orbital component, $|\vec r\uo(t)|^2 \gg |\vec r\us(t)|^2$.
Hence, $|\Delta n\ux| \propto 1/|\vec r\uo(t)|$, and the coordinate resolution is largely determined by the orbital motion.
If, however, the fitting of the orbital motion described in Section~\ref{sec:least-squares-linear} is effective, the magnitude of the transformed supersky metric elements, e.g.\ $g\uss\pr(n\ux, n\ux)$, will instead be proportional to magnitude of the unfitted diurnal motion, $|\vec r\us(t)|^2$, and hence the coordinate resolution $|\Delta n\ux| \propto 1/|\vec r\us(t)|$ will be coarser by a factor $\sim |\vec r\uo(t)|/|\vec r\us(t)|$.

\subsection{Diagonalizing the metric}\label{sec:diagonalizing-metric}

Now that we have a transformed supersky metric, $\mat g\uss\pr$, with a greatly reduced condition number, we can reliably apply two further transformations to diagonalize it, and arrive at the metric $\mat g\uss\dg$ in the form shown in Eq.~\eqref{eq:g-ss-diagonal}.

The first transformation removes the elements of $\mat g\uss\pr$ which couple the sky coordinates $\vec n$ to the frequency and spindown coordinates $\ndot f{}\pr$.
We write the matrix $\mat g\uss\pr$ as a block matrix of the form
\begin{equation}
\label{eq:g-ss-pr-block}
\mat g\uss\pr = \begin{pmatrix}
\mat g\pr_{nn} & \mat g\pr_{nf} \\
\mat g\pr_{nf}{}\trsp & \mat g\pr_{ff}
\end{pmatrix} \,, \\
\end{equation}
where $\mat g\pr_{nn}$, $\mat g\pr_{nf}$, $\mat g\pr_{ff}$ are matrices with elements $g\uss\pr(n_i, n_j)$, $g\uss\pr(n_i, \ndot f{}\pr)$, and $g\uss\pr(\ndot f{}\pr, \ndot[s']f{}\pr)$ respectively.
We then require a coordinate transformation which sets $\mat g\pr_{nf} = \mat 0$.
This condition is satisfied by the coordinates $(\vec n, \ndot \nu)$, where
\begin{align}
\label{eq:reduced-ssky-nu-def}
\ndot \nu &= \ndot f + \vec\Delta^s \cdot \vec n \,, \\
\label{eq:reduced-ssky-Delta-def}
\vec\Delta^s &= \vec\Gamma^s + \mat g\pr_{ff}{}^{-1} \mat g\pr_{nf}{}\trsp \,.
\end{align}
The vectors $\Delta^s$ introduce additional shifts in frequency and spindown which are linear in the sky position $\vec n$.
The metric in these coordinates is given by
\begin{equation}
\label{eq:g-ss-ddg-block}
\mat g\uss\ddg = \begin{pmatrix}
\mat g\ddg_{nn} & \mat 0 \\
\mat 0 & \mat g\pr_{ff}
\end{pmatrix} \,, \\
\end{equation}
where
\begin{equation}
\label{eq:g-ss-ddg-nn}
\mat g\ddg_{nn} = \mat g\pr_{nn} - \mat g\pr_{nf} \mat g\pr_{ff}{}^{-1} \mat g\pr_{nf}{}\trsp \,. \\
\end{equation}
The right-hand side of Eq.~\eqref{eq:g-ss-ddg-nn} is the Schur complement of the block form of $\mat g\uss\pr$ given in Eq.~\eqref{eq:g-ss-pr-block}.
Figure~\ref{fig:condition_nums} plots the condition number of the diagonal-rescaled $\matbar g\uss\ddg$, which for $T \lesssim 10$~days is reduced, relative to $\matbar g\uss\pr$, by a few orders of magnitude.

The second transformation diagonalizes the sky coordinate block $\mat g\ddg_{nn}$ of the matrix $\mat g\uss\ddg$.
We introduce \emph{aligned supersky} coordinates $(n\ua, n\ub, n\uc)$, such that the metric takes the form [c.f. Eq.~\eqref{eq:g-ss-diagonal}]:
\begin{equation}
\label{eq:g-ss-dg-block}
\mat g\uss\dg = \begin{pmatrix}
g\dg_{n\ua,n\ua} & 0 & 0 \\
0 & g\dg_{n\ub,n\ub} & 0 & \mat 0 \\
0 & 0 & g\dg_{n\uc,n\uc} \\
& \mat 0 & & \mat g\pr_{ff}
\end{pmatrix} \,.
\end{equation}
This is achieved by eigendecomposing $\mat g\ddg_{nn}$ as
\begin{equation}
\label{eq:g-ss-ddg-nn-decomp}
\mat g\ddg_{nn} = \mat Q\dg \mat \Lambda\dg \mat Q\dg{}\trsp \,,
\end{equation}
where the elements of the diagonal matrix $\mat \Lambda\dg$ are the eigenvalues $g\dg_{n\ua,n\ua} \ge g\dg_{n\ub,n\ub} \ge g\dg_{n\uc,n\uc}$, and the columns of $\mat Q$ are the corresponding eigenvectors $\vec Q_{n\ua}$, $\vec Q_{n\ub}$, and $\vec Q_{n\uc}$.
The aligned supersky coordinates are then defined by
\begin{align}
\label{eqs:reduced-ssky-nabc-def}
n\ua &= \vec Q_{n\ua} \cdot \vec n \,, &
n\ub &= \vec Q_{n\ub} \cdot \vec n \,, &
n\uc &= \vec Q_{n\uc} \cdot \vec n \,.
\end{align}

In summary, we have found the diagonal metric $\mat g\uss\dg$ of Eq.~\eqref{eq:g-ss-diagonal}, and associated coordinates $(n\ua, n\ub, n\uc, \ndot \nu)$, by performing the following steps:
\begin{enumerate}[(i)]

\item Compute the metric $\mat g\uss\pr$ and vectors $\vec\Gamma^s$, by following the procedure described in Section~\ref{sec:modell-orbit-moti}.

\item Partition $\mat g\uss\pr$ into a block matrix, as given by Eq.~\eqref{eq:g-ss-pr-block}.

\item Compute the matrix $\mat g\ddg_{nn}$, given by Eq.~\eqref{eq:g-ss-ddg-nn}.

\item Eigendecompose $\mat g\ddg_{nn}$ into eigenvectors $\vec Q_{n\ua}$, $\vec Q_{n\ub}$, $\vec Q_{n\uc}$, and eigenvalues $g\dg_{n\ua,n\ua}$, $g\dg_{n\ub,n\ub}$, $g\dg_{n\uc,n\uc}$, following Eq.~\eqref{eq:g-ss-ddg-nn-decomp}.

\item Compute the vectors $\vec\Delta^s$, given by Eq.~\eqref{eq:reduced-ssky-Delta-def}.

\end{enumerate}
The metric $\mat g\uss\dg$ is then given by Eq.~\eqref{eq:g-ss-dg-block}, the sky coordinates $(n\ua, n\ub, n\uc)$ by Eqs.~\eqref{eqs:reduced-ssky-nabc-def}, and the frequency and spindown coordinates $\ndot \nu$ by Eq.~\eqref{eq:reduced-ssky-nu-def}.
Figure~\ref{fig:condition_nums} plots the condition number of $\mat g\uss\dg$, which after diagonal rescaling is reduced to $\sim 1$.

It should be noted that the transformation from $\mat g\uss$ to $\mat g\uss\dg$ is invertable, and no approximations or assumptions are made in its derivation.
Thus, the mismatch predicted by $\mat g\uss\dg$ will be identical to that predicted by $\mat g\uss$.

\subsection{Reducing the sky dimensionality}\label{sec:reduc-sky-dimens}

\begin{figure}
\centering
\includegraphics[width=\linewidth]{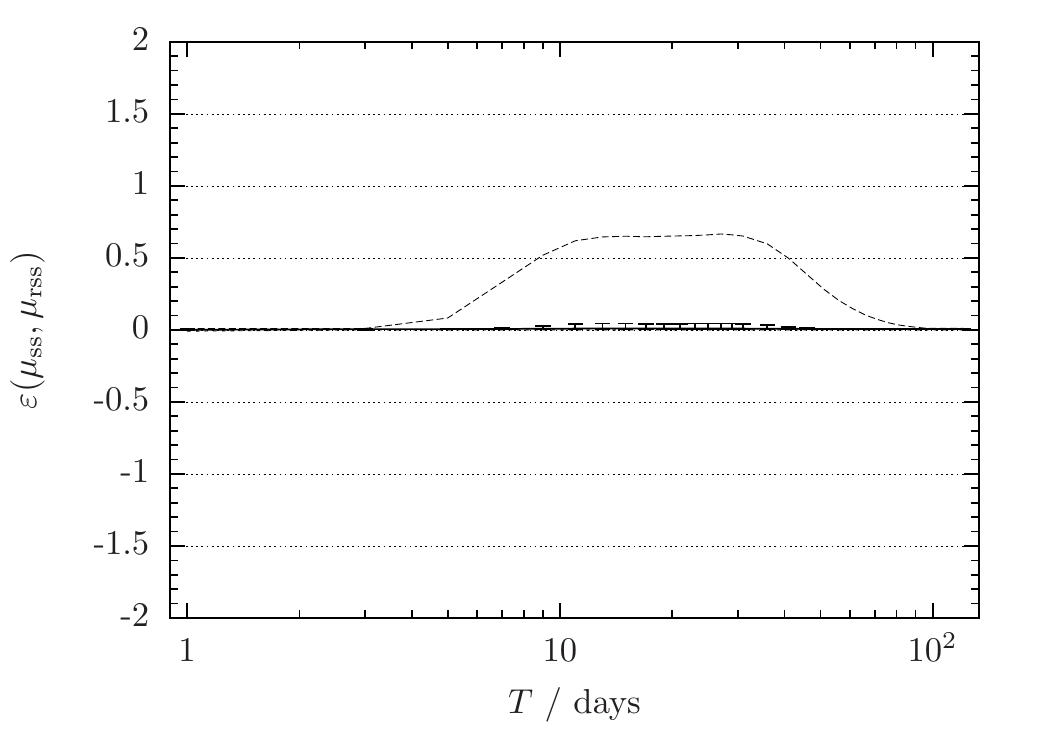}
\caption{\label{fig:mu_re_f1dot_H1_ss_rss_mu0p2}
Relative errors $\relerr{\mu\uss}{\mu\urss}$ between mismatches predicted by the supersky metric, $\mu\uss$, and reduced supersky metric, $\mu\urss$, as a function of $T$, for $\mu\uss \le 0.2$.
Plotted are the median (solid line), the 25th--75th percentile range (error bars), and the 2.5th (dotted line) and 97.5th (dashed line) percentiles.
Only first spindown is used.
}
\end{figure}

\begin{figure*}
\centering
\includegraphics[width=\linewidth]{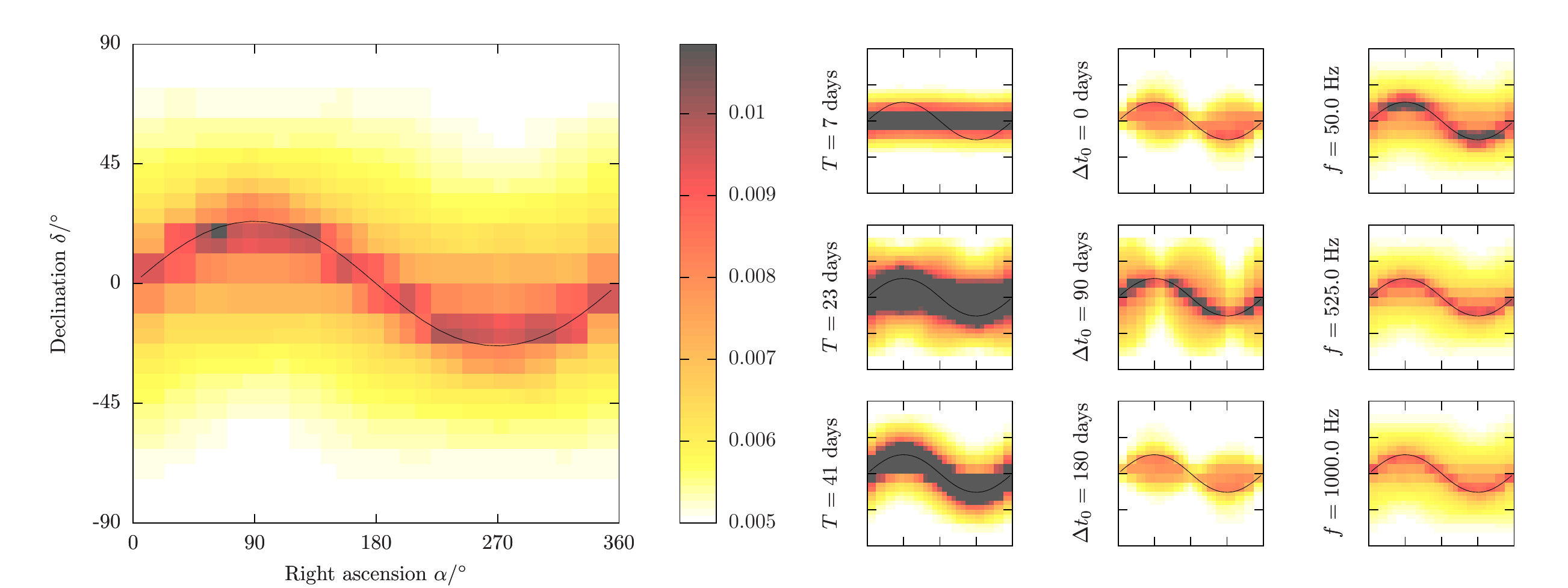}
\caption{\label{fig:mu_sm_f1dot_H1_ss_rss_mu0p2}
Median magnitude of the relative error $|\relerr{\mu\uss}{\mu\urss}|$, as a function of $\alpha$ and $\delta$, for $\mu\uss \le 0.2$.
The ecliptic equator is over-plotted in black.
Left: median $|\relerr{\mu\uss}{\mu\urss}|$ over time-spans $1 \le T \le 121$~days, and over all simulation reference times $\Delta t_0$ and maximum frequencies $f\umax$; see Appendix~\ref{sec:numer-simul}.
Right: median $|\relerr{\mu\uss}{\mu\urss}|$ at fixed values of $T$, $\Delta t_0$, and $f\umax$; axis ranges and color values are the same as for the left-hand-side plot.
Only first spindown is used.
}
\end{figure*}

\begin{figure}
\centering
\subfloat[]{\includegraphics[width=0.5\linewidth]{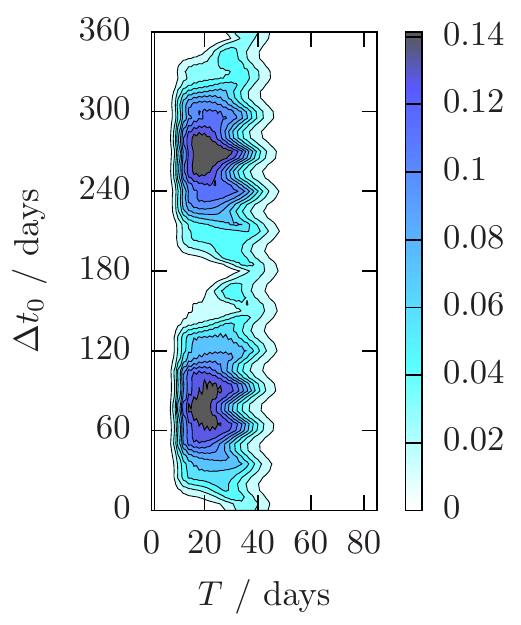}\label{fig:eigrat_f1dot_ts_rt}}
\subfloat[]{\includegraphics[width=0.5\linewidth]{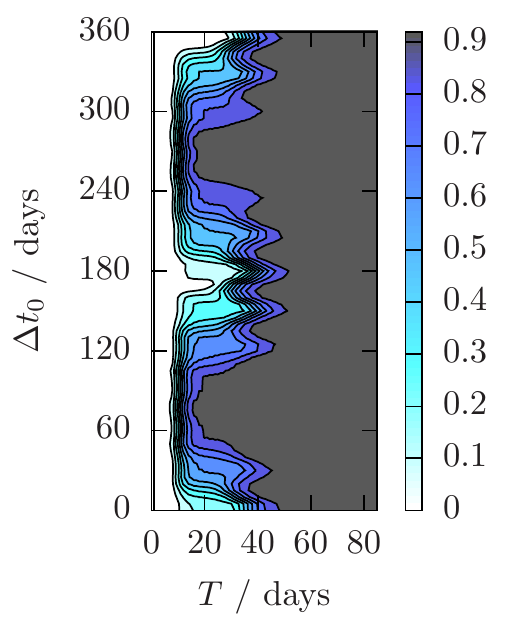}\label{fig:orient_f1dot_ts_rt}}
\caption{\label{fig:f1dot_ts_rt}
\protect\subref{fig:eigrat_f1dot_ts_rt} Ratio of diagonalized supersky eigenvalues $R$, and \protect\subref{fig:orient_f1dot_ts_rt} orientation of $a$--$b$ plane $\beta$, as functions of $T$ and $\Delta t_0$.
Only first spindown is used.
}
\end{figure}

As described at the beginning of Section~\ref{sec:reduced-supersky}, the reduced supersky metric, $\mat g\urss$, takes the diagonalized supersky metric $\mat g\uss\dg$, derived in Sections~\ref{sec:modell-orbit-moti} and~\ref{sec:diagonalizing-metric}, and removes the dimension corresponding to $n\uc$.
This is equivalent to projecting the metric onto the $a$--$b$ plane, as illustrated in Figure~\ref{fig:aligned_ssky_ellipses}.
As a consequence, the mismatch predicted by the reduced supersky metric, $\mu\urss$, will always be smaller than (or equal to) that predicted by the untransformed metric, $\mu\uss$, the difference being $\mu\uss - \mu\urss = g\dg_{n\uc, n\uc} (\Delta n\uc)^2 \ge 0$, where $\Delta n\uc$ is a coordinate offset in $n\uc$.
Since $g\dg_{n\uc, n\uc}$ is, by construction, the smallest of the elements of the sky--sky block of $\mat g\uss\dg$, this represents the smallest error that can be achieved by projecting the sky metric from 3 to 2 dimensions.

Figures~\ref{fig:mu_re_f1dot_H1_twoF_rss_mu0p2} and~\ref{fig:mu_re_f1dot_H1_ss_rss_mu0p2} plot the relative errors, $\relerr{\mu\utwoF}{\mu\urss}$ and $\relerr{\mu\uss}{\mu\urss}$ respectively, between mismatches predicted by the reduced and untransformed supersky metrics, and calculated from the $\calF$-statistic.
The performance of the reduced supersky metric, relative to the $\calF$-statistic (Figure~\ref{fig:mu_re_f1dot_H1_twoF_rss_mu0p2}) is similar to that of the untransformed supersky metric (Figure~\ref{fig:mu_re_f1dot_H1_twoF_ss_mu0p2}).
The one noticeable difference is that there are more trials where the reduced supersky metric significantly under-estimates the $\calF$-statistic mismatch [$\relerr{\mu\utwoF}{\mu\urss} > 0$]; for 2.5\% of trials (above dashed line), $\relerr{\mu\utwoF}{\mu\urss} \gtrsim 0.5$ for $T \lesssim 40$~days.
The origin of this difference is evident in Figure~\ref{fig:mu_re_f1dot_H1_ss_rss_mu0p2}; while the median error $\relerr{\mu\uss}{\mu\urss} \sim 0$, and the 25th--75th percentile range is $\lesssim 0.1$, 2.5\% of trials have $\relerr{\mu\uss}{\mu\urss} > 0.5$ for $10 \lesssim T \lesssim 45$~days.

Figure~\ref{fig:mu_sm_f1dot_H1_ss_rss_mu0p2} plots the median error magnitude $|\relerr{\mu\uss}{\mu\urss}|$ as a function of sky position, over the full ranges of simulation parameters, and at fixed values of $T$, $\Delta t_0$, and $f\umax$.
The largest $|\relerr{\mu\uss}{\mu\urss}|$, and the source of the under-estimation of $\mu\uss$ by $\mu\urss$ observed in Figure~\ref{fig:mu_re_f1dot_H1_ss_rss_mu0p2}, occur along the ecliptic equator, when averaged over $T$, and also at $T = 41$~days.
Note, however, that at $T = 7$~days that the error is largest along the \emph{equatorial} equator $\delta = 0$; at $T = 23$~days the error region is transitioning between equatorial and ecliptic equators.
The largest error is also a function of the reference time: compare $\Delta t_0 = 90$~days with $\Delta t_0 = 0$ and 180~days.
This suggests that the source of the under-estimation can occur either along the equatorial or ecliptic equators, depending in $T$, and is also a function of $\Delta t_0$.

Figure~\ref{fig:f1dot_ts_rt} plots two quantities as functions of $T$ and $\Delta t_0$.
The first, plotted in Figure~\ref{fig:eigrat_f1dot_ts_rt}, is the ratio
\begin{equation}
\label{eq:R-ts-rt-def}
R = \frac{ g\dg_{n\uc, n\uc} }{ g\dg_{n\ub, n\ub} }
\end{equation}
of eigenvalues of the diagonalized supersky metric $\mat g\uss\dg$.
This quantity is a proxy for the error introduced by dropping the term $g\dg_{n\uc, n\uc} (\Delta n\uc)^2$ from the reduced supersky mismatch, relative to the size of the next-largest term in the sky--sky mismatch, $g\dg_{n\ub, n\ub} (\Delta n\ub)^2$.
We see that $R$ increases over the period $10 \lesssim T \lesssim 45$, broadly consistent with the rise and fall of the 97.5th percentile line plotted in Figure~\ref{fig:mu_re_f1dot_H1_ss_rss_mu0p2}.
It is also strongly a function of reference time, being largest at $\Delta t_0 \sim 75$ and 275 days, similar to the median $|\relerr{\mu\uss}{\mu\urss}|$ plotted in Figure~\ref{fig:mu_sm_f1dot_H1_ss_rss_mu0p2}.
Note that $\Delta t_0 \sim 75$ and 275~days roughly coincide with the vernal and autumnal equinoxes~\cite{equinoxes}, UT~$\sim 2007$-03-21 ($\Delta t_0 \sim 79$~days) and UT~$\sim 2007$-09-23 ($\Delta t_0 \sim 265$~days) respectively.

The second quantity, plotted in Figure~\ref{fig:orient_f1dot_ts_rt}, is the angle
\begin{equation}
\label{eq:beta-ts-rt-def}
\beta = \frac{ \cos^{-1} \big| \vec z \cdot \vec Q_{n\uc} \big| }{ \epsilon }
\end{equation}
between the equatorial $z$ axis $\vec z$, and the aligned $c$ axis $\vec Q_{n\uc}$, as a fraction of the Earth's inclination angle $\epsilon$.
When $\beta \sim 0$, at $T \lesssim 10$~days, $\vec z \cdot \vec Q_{n\uc} \sim 1$, and the $a$--$b$ plane is aligned with the equatorial $x$--$y$ plane.
Likewise, when $\beta \sim 1$, at $T \gtrsim 45$~days, $\vec z \cdot \vec Q_{n\uc} \sim \cos\epsilon$, and the $a$--$b$ plane is aligned with the ecliptic $X$--$Y$ plane.
In the period $10 \lesssim T \lesssim 45$~days, the aligned supersky coordinates $(n\ua, n\ub, n\uc)$ transition from ``equatorial-like'' to ``ecliptic-like'' coordinates, at a rate dependent on $\Delta t_0$; note that the $\Delta t_0$ where this transition occurs most rapidly correlate with the $\Delta t_0$ where $R$ is largest.

From the results presented in this section, we may deduce the behavior of the reduced supersky metric as a function of $T$.
When $T \lesssim 10$~days ($\lesssim 3\%$ of 1~year), the fitting performed in Section~\ref{sec:modell-orbit-moti} removes the orbital motion of the Earth from the signal phase, leaving only the diurnal motion of the detector.
This motion is in a plane parallel to the equatorial equator, and hence the $a$--$b$ plane is aligned with the equatorial $x$--$y$ plane, and the aligned supersky coordinates resemble equatorial coordinates (Figure~\ref{fig:orient_f1dot_ts_rt}).
The diagonalized supersky metric ellipsoids are highly elongated perpendicular to the plane of motion, i.e.\ in the equatorial $z$ direction (Figure~\ref{fig:residual_ssky_ellipses}), and hence the ratio of eigenvalues $R$ is small (Figure~\ref{fig:eigrat_f1dot_ts_rt}).

As $T$ increases to between 3\% and 12\% of 1~year ($10 \lesssim T \lesssim 45$~days), the change in signal phase due to the Earth's orbital motion no longer closely resembles a change in phase due to frequency evolution, essentially because it can no longer be modeled as circular motion, which can then be Taylor-expanded (see Section~\ref{sec:modell-orbit-moti}).
Consequentially, the modeling of the orbital motion becomes less effective, and the residuals of the fit, $\delta\bar g\uess(\lambda_i, n\uoSigma)$~[Eq.~\eqref{eq:bar-g-ess-fit}] become larger.
The motion of the detector is therefore a combination of diurnal motion in the equatorial $x$--$y$ plane, and the residual, unfitted orbital motion in the ecliptic $X$--$Y$ plane.
The $a$--$b$ plane is oriented between the equatorial $x$--$y$ and ecliptic $X$--$Y$ planes (Figure~\ref{fig:orient_f1dot_ts_rt}); its exact position is determined by the relative contributions of the diurnal and residual orbital motions.
Since the detector motion is no longer 2-dimensional, the diagonalized supersky metric ellipsoids are no longer highly elongated perpendicular to the $a$--$b$ plane, and hence $R$ increases (Figure~\ref{fig:eigrat_f1dot_ts_rt}).
Because the Earth's orbit is elliptical, the effectiveness of the orbital motion modeling depends on the reference time: at the equinoxes, where the Earth--Sun distance is changing the most rapidly, the Earth's orbit is most poorly approximated by a circular motion, and hence the orbital motion modeling is least effective (Figure~\ref{fig:eigrat_f1dot_ts_rt}).

Eventually, for long enough $T$, the orbital motion can no longer be fitted, and the residuals become equal to the full orbital motion, i.e.\ $\delta\bar g\uess(\lambda_i, n\uoSigma) \sim \bar g\uess(\lambda_i, n\uoSigma)$~[Eq.~\eqref{eq:bar-g-ess-fit}].
The detector motion is then dominated by the full orbital motion, which is much larger than the smaller diurnal motion (as noted in Section~\ref{sec:least-squares-linear}), and the $a$--$b$ plane is aligned with the plane of the Earth's orbit, i.e.\ the ecliptic $X$--$Y$ plane.
The diagonalized supersky metric ellipsoids return to being highly elongated perpendicular to the plane of motion, i.e.\ the ecliptic $Z$ direction (Figure~\ref{fig:residual_ssky_ellipses}), and the ratio $R$ is again small (Figure~\ref{fig:eigrat_f1dot_ts_rt}).

\subsection{Second spindown}\label{sec:second-spindown}

\begin{figure}
\centering
\subfloat[]{\includegraphics[width=\linewidth]{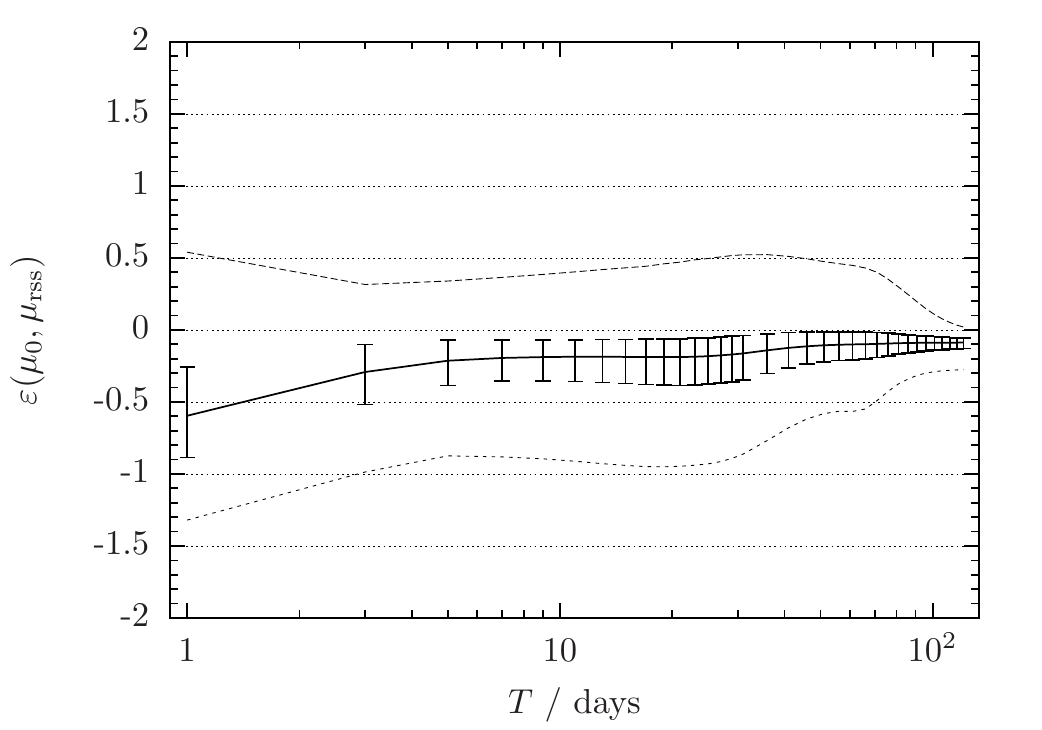}\label{fig:mu_re_f2dot_H1_twoF_rss_mu0p2}}\\
\subfloat[]{\includegraphics[width=\linewidth]{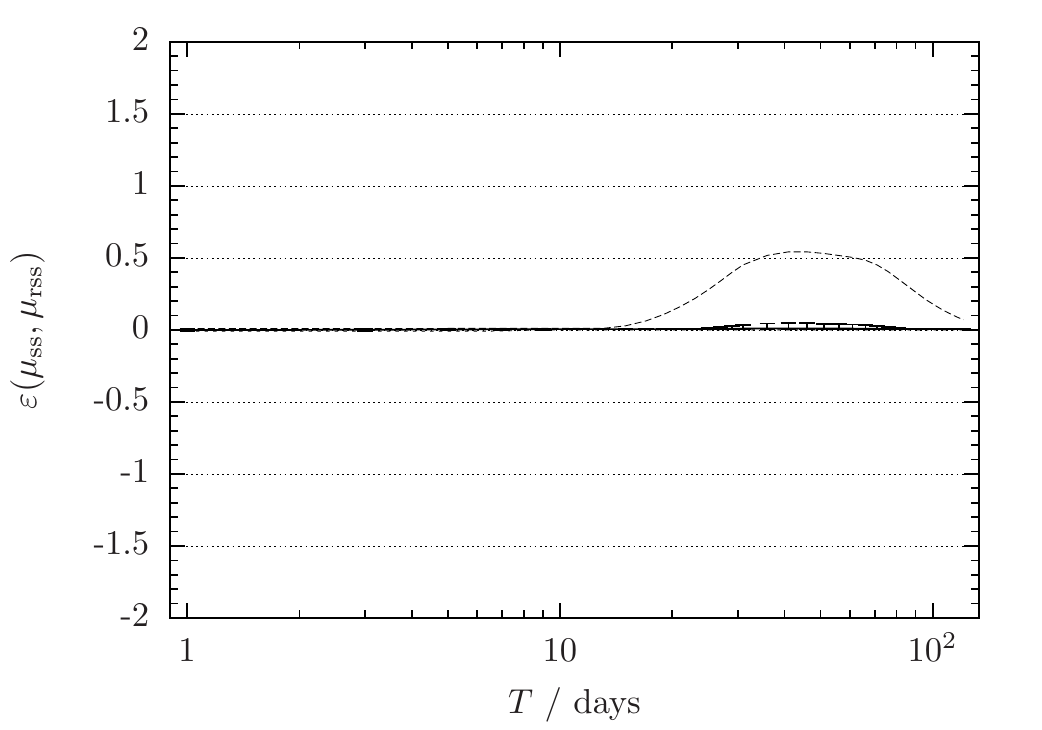}\label{fig:mu_re_f2dot_H1_ss_rss_mu0p2}}
\caption{\label{fig:mu_re_f2dot_H1_rss_mu0p2}
Relative errors \protect\subref{fig:mu_re_f2dot_H1_twoF_rss_mu0p2}~$\relerr{\mu\utwoF}{\mu\urss}$ and \protect\subref{fig:mu_re_f2dot_H1_ss_rss_mu0p2}~$\relerr{\mu\uss}{\mu\urss}$ between mismatches calculated from the $\calF$-statistic, $\mu\utwoF$, and predicted by the supersky metric, $\mu\uss$, and reduced supersky metric, $\mu\urss$, as a function of $T$, for $\mu\utwoF \le 0.2$ and $\mu\uss \le 0.2$ respectively.
Plotted are the median (solid line), the 25th--75th percentile range (error bars), and the 2.5th (dotted line) and 97.5th (dashed line) percentiles.
Both first and second spindowns are used.
}
\end{figure}

\begin{figure}
\centering
\subfloat[]{\includegraphics[width=0.5\linewidth]{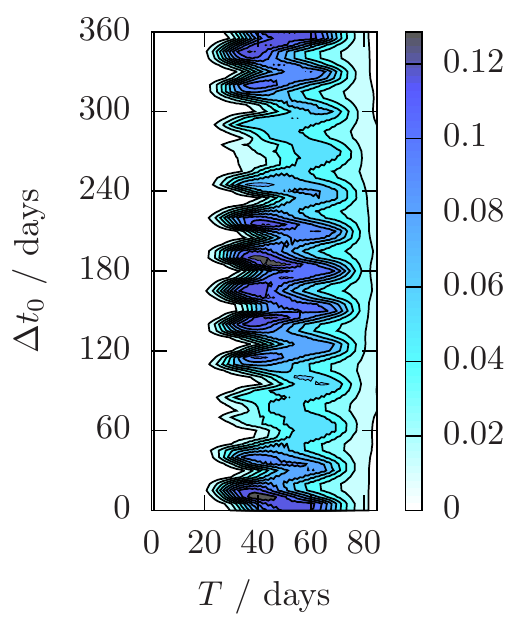}\label{fig:eigrat_f2dot_ts_rt}}
\subfloat[]{\includegraphics[width=0.5\linewidth]{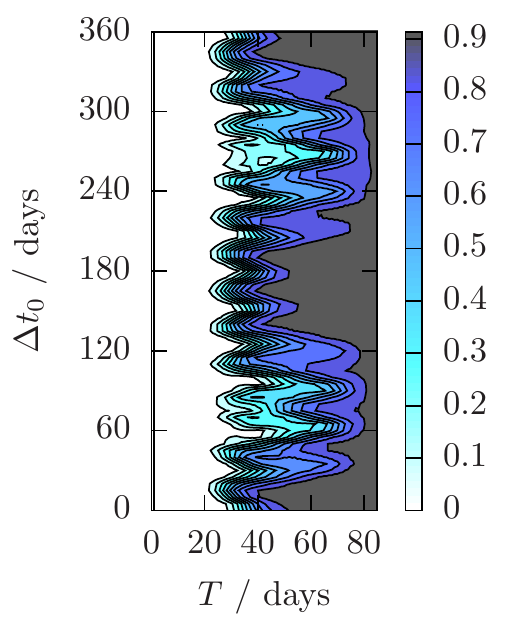}\label{fig:orient_f2dot_ts_rt}}
\caption{\label{fig:f2dot_ts_rt}
\protect\subref{fig:eigrat_f2dot_ts_rt} Ratio of diagonalized supersky eigenvalues $R$, and \protect\subref{fig:orient_f2dot_ts_rt} orientation of $a$--$b$ plane $\beta$, as functions of $T$ and $\Delta t_0$.
Both first and second spindowns are used.
}
\end{figure}

The results presented so far in this paper, with the exception of Figure~\ref{fig:smallest_eigval_err}, have considered parameter-space metrics which include only the first spindown, $\dot f$, in the gravitational-wave pulsar signal model.
In this section, we briefly consider the effect of adding the second spindown, $\ddot f$, to the reduced supersky metric derived in Section~\ref{sec:reduced-supersky}.
Second spindown is important for some potential gravitational-wave pulsars, such as young neutron stars in supernova remnants~\cite{Wette.etal.2008a} and at the Galactic center~\cite{Aasi.etal.2013b}.

Figure~\ref{fig:mu_re_f2dot_H1_rss_mu0p2} plots the relative errors $\relerr{\mu\utwoF}{\mu\urss}$ and $\relerr{\mu\uss}{\mu\urss}$ between mismatches predicted by the reduced and untransformed supersky metrics respectively, and calculated from the $\calF$-statistic, where both first and second spindowns are included in the metrics.
The median errors are similar to those plotted in Figures~\ref{fig:mu_re_f1dot_H1_twoF_rss_mu0p2} and~\ref{fig:mu_re_f1dot_H1_ss_rss_mu0p2}, which include only first spindown.
In Figure~\ref{fig:mu_re_f2dot_H1_twoF_rss_mu0p2}, however, we observe a larger number of trials outside of the 2.5th--97.5th percentile range (dotted to dashed lines) at longer $T$, relative to Figure~\ref{fig:mu_re_f1dot_H1_twoF_rss_mu0p2}.
In Figure~\ref{fig:mu_re_f2dot_H1_ss_rss_mu0p2}, the rise and fall of the 97.5th percentile line spans $30 \lesssim T \lesssim 70$~days, whereas in Figure~\ref{fig:mu_re_f1dot_H1_ss_rss_mu0p2} it spans $10 \lesssim T \lesssim 45$~days.

Figure~\ref{fig:f2dot_ts_rt} plots $R$~[Eq.~\eqref{eq:R-ts-rt-def}] and $\beta$~[Eq.~\eqref{eq:beta-ts-rt-def}] as functions of $T$ and $\Delta t_0$.
Whereas in Figure~\ref{fig:eigrat_f1dot_ts_rt} the ratio $R$ is largest at the equinoxes, in Figure~\ref{fig:eigrat_f2dot_ts_rt} it is largest at the solstices.
Comparing Figure~\ref{fig:orient_f2dot_ts_rt} to Figure~\ref{fig:orient_f1dot_ts_rt}, we see that, with second spindown included, the orbital motion fitting is effective for longer $T$, and that the transition of the aligned supersky coordinates from equatorial-like to ecliptic-like occurs over the later period $20 \lesssim T \lesssim 75$~days, consistent with the rise and fall of the 97.5th percentile line in Figure~\ref{fig:mu_re_f2dot_H1_ss_rss_mu0p2}.
In short, including second spindown changes the behavior of the reduced supersky metric as a function of $T$.

\section{Discussion}\label{sec:discussion}

\begin{figure}
\centering
\includegraphics[width=\linewidth]{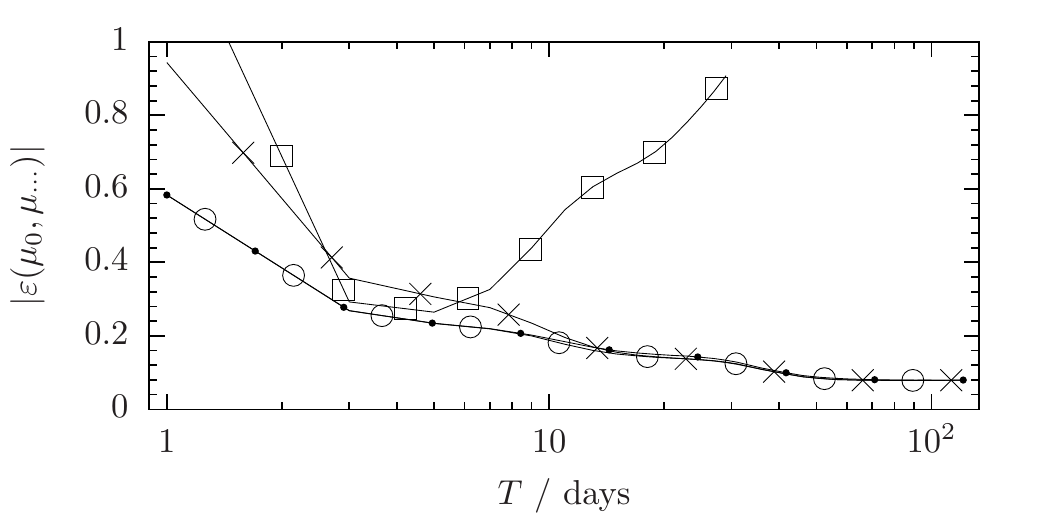}
\caption{\label{fig:compare_mismatch_relerr}
Median absolute relative errors, as functions of $T$, between mismatches calculated from the $\calF$-statistic and predicted by: the supersky metric, $|\relerr{\mu\utwoF}{\mu\uss}|$ (points); the reduced supersky metric, $|\relerr{\mu\utwoF}{\mu\urss}|$ (circles); the linear phase model I metric, $|\relerr{\mu\utwoF}{\mu\usslpI}|$ (crosses); and the global correlation metric, $|\relerr{\mu\utwoF}{\mu\ugct}|$ (squares); for $\mu\utwoF \le 0.2$.
Only first spindown is used.
}
\end{figure}

\begin{figure}
\centering
\includegraphics[width=\linewidth]{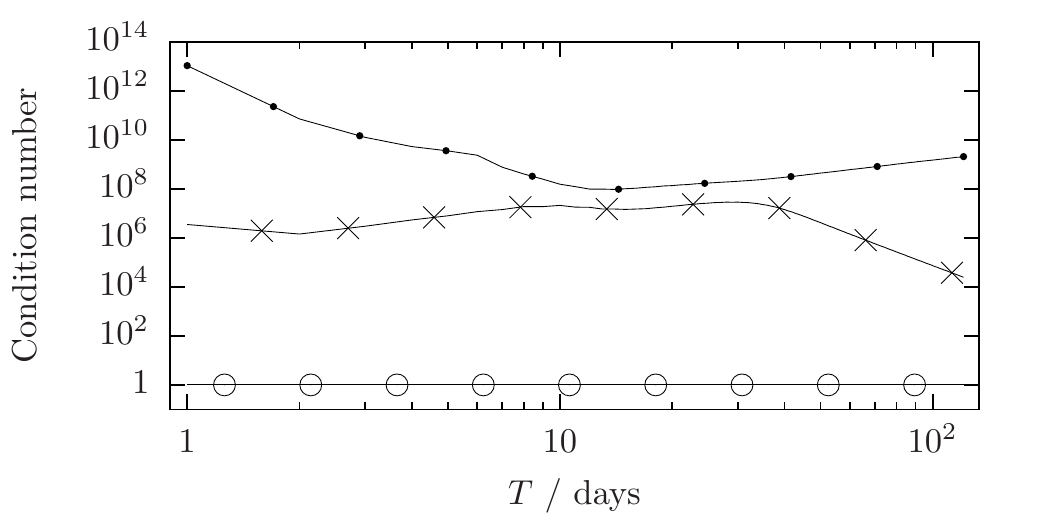}
\caption{\label{fig:compare_condition_nums}
Mean condition numbers, as functions of $T$, of the diagonally-rescaled supersky (points), reduced supersky (circles), and linear phase model I (crosses) metrics, over all simulation reference times $\Delta t_0$; see Appendix~\ref{sec:numer-simul}.
Only first spindown is used.
}
\end{figure}

In this paper we have presented a new explicitly flat metric approximation, the reduced supersky metric $\mat g\urss$, with associated coordinates $(n\ua,\,n\ub,\,\nu,\,\dot \nu, \dots)$, for performing all-sky coherent searches for gravitational-wave pulsars.
Unlike previous work, the reduced supersky metric places no limitation on the timespan $T$ that can be coherently analyzed.
In addition, compared to previous metrics, the reduced supersky metric is well-conditioned.

Figure~\ref{fig:compare_mismatch_relerr} compares the median absolute relative errors in predicted mismatch of the supersky, reduced supersky, linear phase model I, and global correlation metrics.
The supersky metrics perform better at predicting the $\calF$-statistic mismatch than both the linear phase model I and global correlation metrics, for $T \lesssim 10$~days, and are similar to linear phase model I for larger $T$.
Additionally, as shown in Figure~\ref{fig:compare_condition_nums}, the reduced supersky metric has the practical advantage of a much smaller condition number (after diagonal rescaling) than both the supersky and linear phase model I metrics.

Future work will focus on extending this method to the semi-coherent metric, which is required for placing templates in a hierarchical search, where
coherently-analyzed data segments are incoherently combined.
In principle, by allowing $T$ to be increased, the reduced supersky metric can help improve the sensitivity of all-sky hierarchical searches.

Another interesting application of the reduced supersky metric would be in follow-up pipelines for interesting gravitational-wave pulsar candidates~\cite{Shaltev.Prix.2013a}.
Starting with a list of candidates from an initial all-sky search, the follow-up refines the candidates using a series of semi-coherent searches with increasing coherence time $T$ (thereby increasing sensitivity). Finally the follow-up concludes by performing a fully-coherent search of the remaining candidates.
It would be most convenient for such a follow-up pipeline to be able to make use of a parameter-space metric that remains valid for all $T$, without having to switch and/or interpolate between different metric approximants.
Therefore the reduced supersky metric could assist in performing more refining searches, with a smoother increase in $T$, thereby allowing more candidates to be followed up and increasing the chances of detection.

\acknowledgments

We thank Bruce Allen, Badri Krishnan, Gian Mario Manca, Chris Messenger, and Holger Pletsch for helpful discussions.
Numerical simulations were performed on the ATLAS computer cluster of the Max-Planck-Institut f\"ur Gravitationsphysik.
This paper has document numbers AEI-2013-247 and LIGO-P1300155-v3.

\appendix

\section{Numerical simulations}\label{sec:numer-simul}

This appendix details the numerical simulations presented in this paper.

The relative error comparisons presented in Figures~\ref{fig:mu_re_f1dot_H1_twoF}, \ref{fig:mu_sm_f1dot_H1_twoF_gct_mu0p2}, \ref{fig:mu_sm_f1dot_H1_twoF_ss_mu0p2}, \ref{fig:mu_re_f1dot_H1_ssad_mu0p2}, \ref{fig:mu_sm_f1dot_H1_ss_rss_mu0p2}, \ref{fig:mu_re_f2dot_H1_rss_mu0p2}, and~\ref{fig:compare_mismatch_relerr} were produced by Monte-Carlo simulations, as follows.
For each trial, pairs of gravitational-wave signal parameters, given in the reduced supersky coordinates of Section~\ref{sec:reduced-supersky} by $\vec\lambda\pr_1 = ({n\ua}_1, {n\ub}_1, \ndot \nu_1)$ and $\vec\lambda\pr_2 = ({n\ua}_2, {n\ub}_2, \ndot \nu_2)$ were randomly chosen, such that the mismatch $\mu\urss$ between them, computed using the reduced supersky metric $\mat g\urss$, is uniformly distributed within $0 \le \mu\urss \le 0.6$.
To achieve this, a random point $\vec p$ is chosen uniformly within the $(3 + s\umax)$-dimensional unit sphere, and used to compute $\vec\lambda\pr_2 - \vec\lambda\pr_1 = \sqrt{0.6} \, \mat G\urss^{-1} \vec p$, where $\mat G\urss$ is the Cholesky decomposition of $\mat g\urss$ (i.e.\ $\mat G\urss\trsp \mat G\urss = \mat g\urss$).
The reason for using the reduced supersky metric here is that a well-conditioned metric is required to compute the Cholesky decomposition, which in turn
is needed for sampling uniformly with respect to the metric.
Once the offset $\vec\lambda\pr_2 - \vec\lambda\pr_1$ is determined in this way, we can chose $\vec\lambda\pr_1$ uniformly in $({n\ua}_1, {n\ub}_1)$ over the unit disc, and $\ndot \nu_1$ uniformly in frequency and spindown(s).
We then compute ${n\uc}_1 = \sqrt{1 - {n\ua}_1^2 - {n\ub}_1^2}$, and transform the coordinates $({n\ua}_1, {n\ub}_1, {n\uc}_1, \ndot \nu_1)$ back to the untransformed supersky coordinates $\vec\lambda_1 = (\vec n_1, \ndot f_1)$; similarly for ${n\uc}_2$ and $\vec\lambda_2$.
These coordinates are then used to compute the untransformed supersky mismatch $\mu\uss$ and, with additional coordinate transformations, the linear phase model mismatches $\mu\usslpI$ and $\mu\usslpII$, the global correlation mismatch $\mu\ugct$, and the mismatch in $\alpha$--$\delta$ coordinates $\mu\ussad$.
The global correlation metric was computed from the equations in~\cite{Pletsch.2010a}; all other metrics were computed numerically using the software package \textsf{LALPulsar}~\footnote{
Available from \url{https://www.lsc-group.phys.uwm.edu/daswg/projects/lalsuite.html}
}, using the standard JPL ephemerides for the Earth's orbital motion.
Finally, the $\calF$-statistic mismatch $\mu\utwoF$ is calculated as described in Section~\ref{sec:metric}, using the implementation of the $\calF$-statistic in \textsf{LALPulsar}.
Template banks are usually generated with mismatches of the order $\mu\umax \sim 0.2$--0.3~\cite[e.g.][]{Abadie.etal.2010b,Aasi.etal.2013a}; in this paper we generally restrict the sampled mismatches to $\mu\utwoF \le 0.2$.

The above procedure is parametrized by the coherent time-span $T$ and reference time $\Delta t_0 = t_0 - \text{UTC}$~2007-01-01~00:03:06 of the simulated data, and the maximum frequency $f\umax$ of the simulated signals.
We perform the simulations at: 34 values of $T$, from 1~to 31~days in steps of 2~days, and from 36~to 121~days in steps of 5~days; 25 values of $\Delta t_0$, from 0~to 360~days in steps of 15~days; and 5 values of $f\umax \in \{50, 287.5, 525, 762.5, 1000\}$~Hz.
In total, $\sim 10^8$~trials are performed.
When plotted, however, the trials are filtered by restrictions on the mismatches, e.g.\ $\mu\utwoF \le 0.2$ in Figure~\ref{fig:mu_re_f1dot_H1_twoF_sslpI_mu0p2}.
The data from the simulations is also used to generate Figures~\ref{fig:offset_maps} and~\ref{fig:mu_twoF_vs_mu_ss}.

Figures~\ref{fig:condition_nums}, \ref{fig:smallest_eigval_err}, \ref{fig:f1dot_ts_rt}, \ref{fig:f2dot_ts_rt}, and~\ref{fig:compare_condition_nums} were created by computing the various super sky metrics $\mat g\uss$, $\mat g\urss$, etc., using \textsf{LALPulsar}, at: 121 values of $T$, from 1~to 121~days in steps of 1~day; 73 values of $\Delta t_0$, from 0~to 360~days in steps of 5~days; and at $f\umax = 1000$~Hz.

\bibliography{CoherentSkyMetricPaper}

\end{document}